\documentclass[superscriptaddress,groupedaddress,nofootnoteinbib,11pt]{article}
\pdfoutput=1 % if your are submitting a pdflatex (i.e. if you have
\usepackage{jheppub} % for details on the use of the package, please
\usepackage{amsmath, amsfonts, amsthm, amssymb, graphicx, color, hyperref}%, slashed, braket}

\usepackage{float, subfig}
\usepackage{pstool}
\usepackage[utf8]{inputenc}
\usepackage[normalem]{ulem}
\usepackage[english]{babel}
\usepackage{blkarray}
\usepackage{dsserif}
\usepackage{mathtools}
\usepackage{makecell}
\usepackage{multicol}
\usepackage{cleveref}
\usepackage{url}
\usepackage{hyperref}

\allowdisplaybreaks[3]
\def \bal#1\eal  {\begin{align} #1 \end{align}}
\def\({\left(}
\def\){\right)}
\def\[{\left[}
\def\]{\right]}
\def\<{\langle}
\def\>{\rangle}
\def\d{\mathrm{d}}

\newcommand{\eref}[1]{Eq.~\eqref{#1}}
\newcommand{\f}[2]{\frac{#1}{#2}}
\newcommand{\bim} {\begin{itemize}[noitemsep]}

\newcommand{\eim}{\end{itemize}}
\newcommand{\be} {\begin{equation}}
\newcommand{\ee} {\end{equation}}
\newcommand{\bc}{\begin{center}}
\newcommand{\ec}{\end{center}}

\newcommand{\im}{{\rm Im}}

\newcommand{\mc} {\mathcal}

\newcommand{\gi}{{\gamma}}

\newcommand{\epi}{\epsilon}
\newcommand{\thi}{\theta}

\newcommand{\re}{{\rm Re}}
\usepackage[dvipsnames]{xcolor}
\definecolor{c1}{HTML}{D9A1A7}
\definecolor{c2}{HTML}{A1B4A5}
\definecolor{c3}{HTML}{FCD38F}
\definecolor{c4}{HTML}{309F96}

\title{Primal S-matrix bootstrap with dispersion relations}

\author[a]{Claudia de Rham}
\author[a]{, Andrew J. Tolley}
\author[b,c]{, Zhuo-Hui Wang}
\author[b,c]{and Shuang-Yong Zhou}

\affiliation[a]{Abdus Salam Centre for Theoretical Physics, Imperial College, London, SW7 2AZ, UK}
\affiliation[b]{Interdisciplinary Center for Theoretical Study, University of Science and Technology of China, Hefei, Anhui 230026, China}
\affiliation[c]{Peng Huanwu Center for Fundamental Theory, Hefei, Anhui 230026, China}

\emailAdd{c.de-rham@imperial.ac.uk}
\emailAdd{a.tolley@imperial.ac.uk}
\emailAdd{wzh33@mail.ustc.edu.cn}
\emailAdd{zhoushy@ustc.edu.cn}

\preprint{{\footnotesize Imperial/TP/2025/cdr/3~~~~~~~~ USTC-ICTS/PCFT-25-24}}

\date{\today}

\abstract{  
 We propose a new method for constructing the consistent space of scattering amplitudes by parameterizing the imaginary parts of partial waves and utilizing dispersion relations, crossing symmetry, and full unitarity. Using this framework, we explicitly compute bounds on the leading couplings and examine the Regge behaviors of the constructed amplitudes. The method also readily accommodates spinning bound states, which we use to constrain glueball couplings. By incorporating dispersion relations, our approach inherently satisfies the Froissart-Martin/Jin-Martin bounds or softer high-energy behaviors by construction. This, in turn, allows us to formulate a new class of fractionally subtracted dispersion relations, through which we investigate the sensitivity of coupling bounds to the asymptotic growth rate.}

\begin{document}

\maketitle

\section{Introduction}

The modern revival of the S-matrix bootstrap has seen remarkable progress in recent years \cite{Paulos:2016fap, Paulos:2016but, Paulos:2017fhb, Guerrieri:2018uew, Doroud:2018szp, He:2018uxa, Cordova:2018uop, Homrich:2019cbt, EliasMiro:2019kyf, Correia:2020xtr, Bose:2020shm, Bose:2020cod, Hebbar:2020ukp, Guerrieri:2021ivu, He:2021eqn, Guerrieri:2021tak, Chen:2021pgx, Chen:2022nym, EliasMiro:2022xaa, Haring:2022cyf, Antunes:2023irg, Tourkine:2023xtu, Dersy:2023job, He:2023lyy, Haring:2023zwu, Guerrieri:2023qbg, Eckner:2024ggx, Guerrieri:2024ckc, Bhat:2024agd, Copetti:2024dcz, Guerrieri:2024jkn, Gumus:2024lmj,  He:2025gws, Correia:2025uvc} (see \cite{Kruczenski:2022lot} for a recent review), driven by the development of new analytical frameworks and the application of a variety of numerical techniques. A significant focus has been placed on bootstrapping bounds on low energy coefficients in an effective field theory setup (see, {\it e.g.}, \cite{Adams:2006sv, deRham:2017avq, Bellazzini:2018paj, Zhang:2018shp, Arkani-Hamed:2020blm, Bellazzini:2020cot, Tolley:2020gtv, Caron-Huot:2020cmc, Sinha:2020win, Zhang:2020jyn, Remmen:2020vts,  Chiang:2021ziz, Li:2021lpe, Bern:2021ppb, Caron-Huot:2021rmr, deRham:2022gfe, deRham:2021bll, Albert:2022oes, Haring:2022sdp, Henriksson:2022oeu, CarrilloGonzalez:2023cbf, Berman:2023jys, Albert:2023seb, Berman:2024wyt, Albert:2024yap, Caron-Huot:2024lbf, Wan:2024eto, Chang:2025cxc, Beadle:2025cdx, Cheung:2025krg, Bhat:2025zex} and \cite{deRham:2022hpx} for a review), owing to its broad applicability in contemporary research. Departing from the earlier hope of uniquely determining quantum field theories from basic principles such as unitarity, analyticity \cite{Mandelstam:1958xc, Martin:1965jj}, and crossing symmetry \cite{Bros:1965kbd, Mizera:2021ujs}, the modern approach instead seeks to systematically chart the space of consistent theories, reflecting a broader recognition that this space can be rich and extensive.

Due to its simplicity, a massive scalar field serves as an ideal testing ground for bootstrap methods. In the primal bootstrap approach of \cite{Paulos:2017fhb}, a 2-to-2 scattering amplitude $\mathcal{M}(s,t)$ is analytically continued to a tri-variant complex function, cut along $s, t, u \geq 4m^2$, and subject to the constraint $s + t + u = 4m^2$. By introducing the variables $\rho(s)$, $\rho(t)$, and $\rho(u)$ (see Appendix \ref{sec:preprimal}), the triple-cut complex planes are compactified into a product of three unit disks, with the unitarity cuts mapped to the boundaries of these disks. Then, except for possible isolated poles due to bound states below the threshold, a generic amplitude can be parametrized by a tri-variant Taylor expansion around $\rho(s)=\rho(t)=\rho(u)=0$ in terms of the $\rho$ variables. In this formulation, crossing symmetry is manifest as permutation symmetry among the three $\rho$ variables. Partial wave unitarity, on the other hand, imposes nontrivial constraints that can be implemented along the boundaries of the unit disks. As a result, determining the boundary of the allowed amplitude coefficients reduces to a semi-definite programming problem, which can be efficiently solved by modern solvers such as SDPB \cite{Simmons-Duffin:2015qma}.

While the primal method constructs consistent amplitudes explicitly, the dual bootstrap approach proceeds in parallel by using the dispersion relation to exclude regions of parameter space that are inconsistent with fundamental principles~\cite{Lopez:1975ca, Lopez:1976zs, Guerrieri:2021tak, Guerrieri:2023qbg}. (The positivity bounds program for EFTs \cite{deRham:2022hpx} falls within the category of dual bootstrap.) In this approach, analyticity of the amplitude is encoded in the dispersion relation, which expresses the amplitude as an integral over its discontinuity along the unitarity cut. To facilitate the imposition of unitarity constraints, one can employ the partial wave projection of the dispersion relation—namely, the Roy equations~\cite{Roy:1971tc}. While $su$ crossing symmetry can be imposed directly at the level of the fixed-$t$ dispersion relation, $tu$ crossing symmetry must be imposed separately to eliminate odd-spin partial waves, particularly their real parts. To derive bounds on amplitude coefficients in the dual formulation, one begins by constructing a Lagrangian that incorporates the constraints from the Roy equations and partial wave unitarity (within the domain of validity of the Roy equations) via appropriate Lagrange multipliers; Then, by eliminating the partial wave amplitudes ({\it i.e.}, the primal variables, which appear linearly in the Lagrangian) through their equations of motion, one arrives at a semi-definite programming problem formulated entirely in terms of the dual variables \cite{Lopez:1976zs, Guerrieri:2021tak}.  

In this paper, we develop a primal bootstrap method that nonetheless makes essential use of the (fixed-$t$) dispersion relation. By leveraging the dispersion relation, we construct the amplitude space in a manner that is arguably more directly tied to physical processes. Specifically, we decompose the amplitude into partial waves, and parametrize only the imaginary parts of the partial wave amplitudes in the physical region, namely, along the unitarity cut above the threshold $\mu\ge 4m^2$ ($\mu$ being the integration variable along the cut). The real parts of the partial wave amplitudes are then determined by solving a system of linear equations derived from the dispersion relation evaluated at different values of $t$. In this setup, $tu$ crossing symmetry--and hence full crossing symmetry--can be trivially enforced. More importantly, the dispersion relation can now, in principle, be applied in the bootstrap at arbitrarily large values of $s$. Partial wave unitarity conditions can subsequently be used to set up semi-definite programs to derive bounds on the amplitude coefficients such as the leading ones $g_0$ and $g_2$ (cf.~\eref{g0g2coefs} and Figure \ref{fig:2d_g0g2}).

One advantage of constructing the amplitude using a dispersion relation is that the Froissart-Martin bound and its generalization the Jin-Martin bound \cite{Froissart:1961ux, Martin:1962rt,Jin:1964zza}\footnote{Froissart \cite{Froissart:1961ux} derived a bound on the forward limit amplitude assuming the Mandestam double spectral representation and Martin \cite{Martin:1962rt} generalized this proof to any polynomially bounded amplitude. Jin and Martin further extended this to the positive $t$ region $0\le t<4m^2$ \cite{Jin:1964zza}.} is, in some sense, built in: the very existence of the twice-subtracted dispersion relation relies on this asymptotic behavior. In contrast, the primal method formulated in terms of $\rho$ variables builds the amplitude by expanding around a low-energy point, making it less transparent how high-energy behavior can be captured. To assess how the two primal methods differ in realizing the Regge behavior, we explicitly construct a few extremal amplitudes on the boundary of the 2D bound on the leading coefficients $g_0$ and $g_2$ at large $|s|$ and fixed $t$. We find that the amplitude associated with the right-hand kink remains bounded at high energies. In comparison, the amplitude near the left-hand kink diverges as $|s|$ increases. An analysis of the corresponding partial wave spectra reveals that high-spin contributions become increasingly significant as the left kink is approached from below. We find that the most divergent amplitudes grow at least as fast as $s^{1.7 \sim 1.8}$ in both methods, which occurs near the $t$-channel threshold singularity at $t = 4m^2$. Further implications for the UV theory will be explored elsewhere. Generically, the amplitude constructed with our method is slightly greater than that of the previous primal method. Nevertheless, it seems that a large order expansion in the $\rho$-variable primal bootstrap can effectively realize a fast Regge growth at large $|s|$.  
Interestingly, near the left kink, the growth rate exhibits a quasi-linear dependence on $t$, particularly in the positive $t$ region (see Figures \ref{fig:cmp-std-1} and \ref{fig:away_g0min}). Moreover, the region of fastest growth appears to be located away from the left kink.

Thanks to the use of the dispersion relation, our formalism can also easily accommodate bounds states with arbitrary spin below the threshold. For spin-0 bound states, this is also straightforward to achieve using the method that parameterizes the full amplitude. However, extending that approach to higher-spin bound states seems to be challenging, as it requires constructing a crossing-symmetric ansatz with correct residues and asymptotic behavior consistent with the Jin-Martin bound. In contrast, using a fixed-$t$ dispersion relation, bound states with spin can be added systematically in our approach. To further demonstrate its effectiveness, we apply our formalism to glueball scattering. Using the latest Lattice QCD data for the bound-state mass spectrum as input, we derive bounds on the phenomenological couplings of the glueball states (see Figure \ref{fig:glueball}).

The twice-subtracted dispersion relation is usually used because, asymptotically, the Jin-Martin bound remains the strongest bound that is rigorously proven. However, in certain classes of theories, the large-$|s|$ behavior can be significantly softer. Assuming the fixed-$t$ amplitude grows no faster than $|s|^{2r}$, with $r$ a real number, we can take advantage of this by formulating a dispersion relation with a fractional subtraction order. This is achieved by partitioning the unitarity cut above $4m^2$ into several regions, combined with an integration kernel. This partition is also physically motivated, as there are higher multi-particle thresholds above $4m^2$. With this fractionally subtracted setup, we re-compute the bounds on $g_0$ and $g_2$, and find that the upper bound on $g_0$ remains unchanged even as the subtraction order $2r$ approaches zero. In contrast, the lower bound on $g_0$ is only reduced when $2r$ becomes substantially smaller than 1 (see Figure \ref{fig:newdisp_2}).

Note that in the primal bootstrap approach that parametrizes the full amplitude, the multi-particle thresholds are mapped to the unit circles at $|\rho_{s,t,u}|=1$ in the $\rho$ variables, where partial wave unitarity conditions are imposed. However, the $\rho$ expansion does not actually converge at $|\rho_{s,t,u}|=1$ due to the branch-point singularities arising from these multi-particle thresholds. Thus, strictly speaking, partial wave unitarity are more appropriately imposed in a distributional sense \cite{Correia:2020xtr}. In contrast, our method avoids this ambiguity by directly parameterizing the partial waves in the physical region, allowing for an explicit treatment of various multi-particle thresholds. Nevertheless, as we demonstrate in this work, results obtained from both approaches are in good agreement insofar as bounds on theory space are concerned, suggesting that the effects of the branch-point singularities might be minimal.

The rest of the paper is organized as follows. In Section \ref{sec:generalStra}, together with Appendices \ref{sec:thresholdOfa_ell}, \ref{sec:largel} and \ref{sec:cal_int}, we present the main ingredients and steps of our primal bootstrap method, assuming no bound states lie below the threshold. In Section \ref{sec:bound_expansion_coeff}, we apply this method to constrain the leading two Taylor coefficients of the 2-to-2 scattering amplitude and analyze its large-$|s|$ behavior at fixed $t$. Section \ref{sec:spin-0} briefly outlines how to incorporate spin-0 bound states into our formalism and includes a sample bound calculation. In Section \ref{sec:glueball}, we discuss extending the method to cases with higher-spin bound states below the threshold and apply it to the realistic scenario of glueball scattering. Section \ref{sec:fractional} introduces the fractional subtracted dispersion relation for theories with softer Regge behavior and explores how the bounds vary continuously with the subtraction order. We conclude in Section \ref{sec:conclu}. Some example convergence tests are presented in Appendix \ref{sec:cov_test}, and the primal method parametrizing the full amplitude is briefly reviewed in Appendix \ref{sec:preprimal}.

\section{General strategy}
\label{sec:generalStra}

In this section, we will outline our method of constructive bootstrap with the dispersive relation for the case of a massive scalar. For simplicity and illustrative purposes, we assume in this section and the next that the scalar field has no cubic couplings, so that the amplitude exhibits no poles below the threshold. The case with cubic self couplings will be dealt with in Section \ref{sec:spin-0}, and an extended formalism with spinning bound states will be further used to constrain couplings in glueball scattering in Section \ref{sec:glueball}.

\subsection{Dispersion relation}
\label{sec:analyticity_confition}

Consider 2-to-2 scattering for identical scalar particles with mass $m$, and, for now, assume that there is no bound states below the threshold. 
We shall only use the rigorously proven Martin analyticity, which requires that, for fixed-$t$ within $-4m^2<t<4m^2$, the amplitude is analytic on the complex $s$-plane except for the branch cuts at $s \geq 4m^2$ ($s$-channel) and $s \leq -t$ ($u$-channel). The Martin analyticity is actually valid in the complex plane region $|t|<4m^2$. However, due to the analyticity \cite{uniqueAnMath} of the amplitude, it suffices to consider the real 
$t$ axis within this domain. Also by analyticity, the physical amplitudes for different scattering channels correspond to different limits of a single underlying complex amplitude, so we have the real analyticity condition $\mc{M}((s+i\epi)^*,t) = \mc{M}^*(s+i\epi,t)$ (for a real $t$) and the triple crossing symmetry $\mc{M}(s,t) = \mc{M}(s,u)= \mc{M}(t,s)$.
Furthermore, the Regge behavior of the amplitude satisfies the Jin-Martin bound
\begin{equation}
    \lim_{|s|\to \infty}|\mc{M}(s,t)|< |s|^{1+\gi(t)} \text{, for $|t|<4m^2$ and $0<\gi(t)<1$} \,.
\end{equation}
(Theories with Regge growth slower than the Jin-Martin bound will be considered in Section \ref{sec:fractional} with the fractionally subtracted dispersion relation.)
Using these ingredients, the amplitude across the entire complex $s$-plane can be reconstructed from its boundary values on the branch cuts, as realized by the well-known twice-subtracted dispersion relation.

To derive the twice-subtracted dispersion relation, note that the amplitude $\mc{M}(s,t)$ for $-4m^2<t<4m^2$ can be expressed using the Cauchy integration formula as follows:
\begin{equation}\label{disp_origin}
\begin{aligned}
    \mc{M}(s,t) &= \int_{4m^2}^{\infty} \frac{\d\mu}{\pi} \frac{{\rm Disc} \mc{M}(\mu,t)}{\mu-s} + \int_{-\infty}^{-t} \frac{\d\mu}{\pi} \frac{{\rm Disc} \mc{M}(\mu,t)}{\mu-s} +\int_{C^{\infty}}\frac{\d\mu}{2\pi i} \frac{\mc{M}(\mu,t)}{\mu-s}
\\ &=\int_{4m^2}^{\infty} \frac{\d\mu}{\pi} \frac{{\rm Im} \mc{M}(\mu,t)}{\mu-s} + \int_{4m^2}^{\infty} \frac{\d\mu}{\pi} \frac{{\rm Im} \mc{M}(\mu,t)}{\mu-u} +\int_{C^{\infty}}\frac{\d\mu}{2\pi i} \frac{\mc{M}(\mu,t)}{\mu-s}\,,
\end{aligned}
\end{equation}
where in the second line we have used $su$ crossing symmetry $\mc{M}(s,t)=\mc{M}(u,t)$ and real analyticity ${\rm Disc}\mc{M}(s,t) = \frac{1}{2i}\left(\mc{M}(s+i\epsilon,t)-\mc{M}(s-i\epsilon,t)\right)={\rm Im}\mc{M}(s,t)$. (To be pedantic, $\mu$ in the above equations should be understood as $\mu+i\epsilon$.)
The Jin-Martin bound suggests that these integrations may in general diverge due to the large $|s|$ behavior of the amplitude. To proceed, we employ a twice-subtraction for each term on the right hand side, which amounts to a replacement for the denominator of the integrand
\begin{equation}
\label{subtEqu}
   \frac{1}{\mu-z} = \frac{z^2}{(\mu-z)\mu^2} + \frac{z}{\mu^2} + \frac{1}{\mu}\,,
\end{equation}
with $z$ being either $s$ or $u$. The second and third terms in \eref{subtEqu} lead to divergences once substituting into \eref{disp_origin}. However, these divergences must cancel to converge to the amplitude on the left hand side, leading to a finite but undetermined subtraction polynomial remainder\footnote{This is equivalent to using Cauchy's theorem for $\mc{M}(s,t)/[(s-z_0)(u-z_0)]$ for some subtraction point $z_0$.}. The final result is a twice-subtracted dispersion relation that only contain convergent quantities\,\footnote{A possible $b_1(t)s$ term is absent in identical scalar scattering due to $su$ crossing symmetry. In situations where $su$ crossing is not manifest, i.e. $\mc{M}_s(s,t,u)=\mc{M}_u(u,t,s)$, we should include $b_1(t)(s-u)$.}:
\begin{equation}\label{apa0}
    \mc{M}(s,t) = b_0(t)
    +\int_{4m^2}^{\infty} \frac{\d\mu}{\pi} \,  {\rm Im} \mc{M}(\mu,t) \left(\frac{s^2}{\mu^2(\mu-s)} + \frac{u^2}{\mu^2(\mu-u)}\right)\,.
\end{equation}
The amplitude's $st$ crossing symmetry allows us to express $b_0(t)$ in terms of the amplitude at a fixed, analytic, real point $(s_0,t_0)$ of $\mc{M}(s,t)$ \cite{Roy:1971tc, Lopez:1975ca}. To see this, note that 
\begin{equation}\label{apa2}
  \mc{M}(s_0,t_0) =    b_0(t_0) + \int_{4m^2}^{\infty} \frac{\d\mu}{\pi} \, {\rm Im} \mc{M}(\mu,t_0) \left(\frac{s_0^2}{\mu^2(\mu-s_0)} + \frac{(4m^2-s_0-t_0)^2}{\mu^2(\mu-4m^2+s_0+t_0)}\right)\,,
\end{equation}
and $st$ crossing symmetry requires $\mc{M}(t,t_0) = \mc{M}(t_0,t)$ for $-4m^2<t_0<4m^2$, which gives
\begin{equation}\label{apa1}
\begin{aligned}
    b_0(t_0) + \int_{4m^2}^{\infty} \frac{\d\mu}{\pi} \, {\rm Im} \mc{M}(\mu,t_0) \left(\frac{t^2}{\mu^2(\mu-t)} + \frac{(4m^2-t-t_0)^2}{\mu^2(\mu-4m^2+t+t_0)}\right)&\\ =b_0(t) + \int_{4m^2}^{\infty} \frac{\d\mu}{\pi} \, {\rm Im} \mc{M}(\mu,t) \left(\frac{t_0^2}{\mu^2(\mu-t_0)} + \frac{(4m^2-t-t_0)^2}{\mu^2(\mu-4m^2+t+t_0)}\right) \,.
\end{aligned}
\end{equation}
Substituting Eqs.~\eqref{apa2} and \eqref{apa1} into \eref{apa0}, we immediately get
\begin{equation}
    \mc{M}(s,t) = \mc{M}(s_0,t_0) + \frac{1}{\pi}\int_{4m^2}^{\infty}\d\mu \left({\rm Im}\mc{M}(\mu,t) K^{\mu,t_0}_{s,t} + {\rm Im}\mc{M}(\mu,t_0) K^{\mu,s_0}_{t,t_0}\right)\,, \label{disp}
\end{equation}
where the integration kernel $K^{\mu,t_0}_{s,t}$ is defined as 
\begin{equation}
\label{eq:KernelK}
    K^{\mu,t_0}_{s,t} = \frac{1}{\mu-s} + \frac{1}{\mu-4m^2+s+t}-\frac{1}{\mu-t_0}-\frac{1}{\mu-4m^2 +t +t_0}\,.
\end{equation}
We shall choose $s_0$ and $t_0$ to be real, positive and within the Mandelstam triangle: $s_0, t_0, 4m^2-s_0 - t_0 < 4m^2$, which ensures that $\mathcal{M}(s_0, t_0)$ is real.  In this paper we will choose $s_0=t_0=4m^2/3$ except in the Section \ref{sec:glueball} where $s_0=2m^2,~t_0=0$.

In \eref{disp}, $\mu$ and $s$ should be understood as $\mu+i\epsilon$ and $s+i\delta$ respectively, with $0<\epsilon<\delta\ll 1$, as the discontinuity of $\mu$ can be taken first to derive the dispersion relation. With this understanding, using the identity $\frac{1}{x\pm i \epsilon} = \mc{P}\frac{1}{x} \mp i\pi \delta(x)$ for the integration kernel, it is easy to see that the imaginary part of \eref{disp} is trivially satisfied. So the useful information is encoded in the real part, which is given by
\begin{equation}
    \begin{aligned}
    \re \mc{M}(s,t) =  \mc{M}(s_0,t_0)+ \mc{P} \int_{4m^2}^{\infty}\!\!\frac{\d \mu}{\pi} \left(\im \mc{M}(\mu,t) K^{\mu,t_0}_{s,t}  + \im \mc{M}(\mu,t_0) K^{\mu,s_0}_{t,t_0} \right) \,.
    \label{imtoreal}
\end{aligned}
\end{equation}
{\it This is the dispersion relation we will use for our numerical bootstrap.} Given that the dispersion relation is $su$ crossing symmetric and the amplitude is real analytic, it is sufficient to focus on the $s$-channel branch cut for $\mc{M}(s,t)$. Thus, we can restrict to $s\geq 4m^2$, as well as $-4m^2<t<4m^2$, for \eref{imtoreal}. As the boundary of an analytic function, the amplitude on the branch cuts is highly constrained. The real part of the amplitude on the branch cut is fully determined by its imaginary part and the value of $\mathcal{M}(s_0, t_0)$, which plays a crucial role in our method.

\subsection{Parametrizing the partial waves}
\label{sec:paramPartWaves}

In our method, we parametrize the imaginary parts of the partial wave amplitudes to construct the viable space of amplitudes. Let us first review some general properties of partial waves. Recall that the partial wave expansion of a scalar amplitude in 4D is given by: 
\begin{equation}
    \mc{M}(s,t)  = 16\pi \sum_{{\rm even}\, \ell \geq 0} (2\ell +1) a_\ell (s) P_\ell\left(1+\frac{2t}{s-4m^2}\right)\,, 
\end{equation}
where $P_\ell$'s denote the Legendre polynomials. The summation is only over the even $\ell$'s because the odd $\ell$ partial waves vanish, as a result of $tu$ crossing symmetry for identical scalar scattering $\mc{M}(s,t) = \mc{M}(s,u)$. We are interested in the physical region $s > 4m^2$, in which the fixed-$s$ amplitude is analytic in the region $|t| < 4m^2$ on the complex $t$-plane, according to Martin's analyticity. Consequently, the partial wave expansion converges at least within this region of $s$ and $t$. 

Let us look at the behavior of the partial wave amplitudes $a_\ell(s)$ near the threshold $s=4m^2$. As $s$ approaches $4m^2$, the Legendre polynomial $P_\ell\left(1 + 2t/(s-4m^2)\right)$ scales as $(s - 4m^2)^{-\ell}$. For the dispersive integral in \eref{imtoreal} to remain finite, the imaginary part of the partial wave amplitude on the right hand side must decrease faster than $(s - 4m^2)^{\ell - 1}$ as $s$ approaches $4m^2$ (see Appendix \ref{sec:thresholdOfa_ell}): 
\begin{equation}
\label{aellos4m2}
    \lim_{s\to 4m^2} \frac{\im\, a_{\ell}(s)}{(s-4m^2)^{\ell-1}} = 0 \,.
\end{equation}
When parametrizing $\im\, a_{\ell}(s)$, it is essential to explicitly factor out this behavior near $s=4m^2$. However, $(s - 4m^2)^{\ell}$ by itself does not have a desirable behavior at large $s$. Note that unitarity $|1+i\sqrt{(s-4m^2)/{s}}\, a_{\ell}(s)|\leq 1$ implies that $\sqrt{(s-4m^2)/{s}}\, a_\ell(s)$ must be bounded at large $s$. So we can add an extra factor that goes like $s^{-\ell}$ at large $s$. Many choices are possible, but the large $\ell$ behavior of partial waves gives rise to a natural choice (see Appendix \ref{sec:largel}): 
\begin{equation}\label{factor_cut}
    a_\ell(s)=\left( \frac{s-4m^2}{s+4m^2+4m\sqrt{s}}\right)^\ell \Bigg(\cdots\Bigg) ,
\end{equation}
which significantly improves the numerical evaluations later. 

With these observations, we find that it is convenient to parameterize all possible imaginary parts of the partial wave amplitudes as follows:
\begin{equation}
\label{imaansatz0}
    \im\, a_{\ell}(s) = \sqrt{\frac{s}{s-4m^2}}\left( \frac{s-4m^2}{s+4m^2+4m\sqrt{s}}\right)^\ell\, \sum_{k=0}^{k_{\rm max}} c_{\ell,k} P_k(8m^2/s -1)\,,
\end{equation}
where $P_k$'s are again Legendre polynomials and $c_k$ are arbitrary coefficients. In this ansatz, the choice of $8m^2/s-1$ as the argument for the Legendre polynomial is such that when $s$ goes from $4m^2$ to $+\infty$ the argument falls within the region $[-1,1]$. That is, the infinite functional space has been truncated to an $(k_{\rm max} + 1)$-dimensional subspace, and to approximate a bounded function, we have used the Legendre polynomials. The phase space factor $\sqrt{s/(s - 4m^2)}$ is explicitly included because it is the quantity $\sqrt{(s - 4m^2)/s} \, \operatorname{Im} a_{\ell}(s)$, rather than $\operatorname{Im} a_{\ell}(s)$, that is bounded. (This is similar to adding the ${1}/(\rho-1)$ terms in the primal bootstrap method of \cite{Paulos:2017fhb}.)  However, due to the presence of the factor $(s-4m^2)^\ell$ in \eref{imaansatz0}, this factor is not essential for the $\ell>0$ waves. Note that, although the $(\cdots)^\ell$ overall factor in \eref{imaansatz0} only captures the large spin structure of the partial waves, the deviations from this factor at small $\ell$ can be well accommodated by the $P_k(8m^2/s -1)$ expansion. 

In some circumstances, slightly stronger conditions can be imposed on the partial wave amplitudes. For example, it is often required that the scattering lengths \cite{Correia:2020xtr}
\begin{equation}
    l_{\ell}:=\lim_{s\to 4m^2} \frac{a_{\ell}(s)}{(s/4m^2-1)^\ell}  \,,
\end{equation}
are finite for $\ell\geq 2$. Although not explicitly considered in this paper, in this scenario, we can instead adopt the following ansatz 
\begin{equation}
\begin{aligned}
    \im\, a_{0}(s) &= \sqrt{\frac{s}{s-4m^2}}\sum_{k=0}^{k_{\rm max}} c_{0,k} P_k(8m^2/s -1)\,,\\
    \im\, a_{\ell}(s) &= \left( \frac{s-4m^2}{s+4m^2+4m\sqrt{s}}\right)^\ell \sum_{k=0}^{k_{\rm max}} c_{\ell,k} P_k(8m^2/s -1)\text{, for $\ell \geq 2$}\,.
\end{aligned}
\end{equation}

\subsection{The method}
\label{sec:method}

The twice-subtracted dispersion relation \eqref{imtoreal} allows us to construct the fixed-$t$ amplitude from $\mathcal{M}(s_0, t_0)$ and $\operatorname{Im} \mathcal{M}(\mu\geq 4m^2, t)$. Thus, to build all possible amplitudes, it is sufficient to find all possible $\mathcal{M}(s_0, t_0)$ and $\operatorname{Im} \mathcal{M}(\mu\geq 4m^2, t)$. With the point $(s_0, t_0)$ chosen within the Mandelstam triangle,  $\mathcal{M}(s_0, t_0)$ is simply a real number as discussed below \eqref{eq:KernelK}. As for $\operatorname{Im} \mathcal{M}(\mu\geq 4m^2, t)$, we can use a partial wave expansion and  parametrize the resulting amplitudes with the ansatz \eqref{imaansatz0}. Further truncating the spins at $\ell_{\rm max}$, we adopt the following ansatz for the $\operatorname{Im} \mathcal{M}(\mu\geq 4m^2, t)$ in the dispersive integral: 
\begin{align}
\label{ansatz_im}
    \im \mc{M}(\mu,t) &= 16\pi\!\! \sum_{\ell\geq 0, {\rm even}}^{\ell_{\rm max}}\sum_{k=0}^{k_{\rm max}}(2\ell +1) \sqrt{\frac{\mu}{\mu-4m^2}}\left( \frac{\mu-4m^2}{\mu+4m^2+4m\sqrt{\mu}}\right)^\ell 
    \\
    &\qquad\qquad \times c_{\ell,k} P_k\left(\frac{8m^2}{\mu} -1\right)P_\ell\left(1+\frac{2t}{\mu-4m^2}\right)\text{, ~~for $-4m^2<t<4m^2$} \,.\nonumber
\end{align}
With this ansatz, the integration over $\mu$ on the right hand side of \eqref{imtoreal} can be explicitly carried out, albeit in general numerically for given $s,t,s_0,t_0$. Note that this $\mu$ integration contains singular points, which requires taking the principal values, as already noted in \eref{imtoreal}. For the numerical results to converge, it is essential to evaluate the integration near the singular points to a high precision, which is time consuming numerically. Efficient schemes, which will be discussed in Appendix~\ref{sec:cal_int}, are needed to accelerate the numerical integration, as various numerical setups in our formalism require testing. However, once a numerical setup is shown to yield convergent results, the $\mu$ integration outcomes can be reused across all relevant computations.

Our strategy is then to impose full partial wave unitarity conditions to construct all viable amplitudes. However, we have only parametrized $\im\, a_{\ell}(\mu)$. To impose full unitarity, we also need the corresponding real parts of the partial wave amplitudes, which can be obtained via the dispersion relation \eqref{imtoreal}.
To this end, we also expand the left hand side of \eqref{imtoreal} with partial waves 
\begin{equation}
\label{extract_real}
    \re \mc{M}(s,t) =  16\pi\sum_{{\rm even}\,\ell\geq 0}^{\ell_{\rm max}}(2\ell +1) \re\, a_{\ell} (s) P_\ell\left(1+\frac{2t}{s-4m^2}\right) \,.
\end{equation}
For a given ansatz of $\im\, a_{\ell}(\mu)$, the right-hand side of \eqref{imtoreal} is known. So, to obtain $\re\, a_{\ell} (s)$ for each fixed $s$, we can choose a number of different $t$'s for \eqref{imtoreal}, which form a set of linear equations that can be used to solve for $\re\, a_{\ell}(s)$:
\be
\label{Re_a_ellSol}
\begin{aligned}
\qquad t&=t_{\rm min}:& \quad (...)\re\, a_{0}(s) + (...)\re\, a_{2}(s) +  (...)\re\, a_{4}(s) +\cdots  &=(...)
  \\
\qquad t&=t_1: &\quad (...)\re\, a_{0}(s) + (...)\re\, a_{2}(s) +  (...)\re\, a_{4}(s)  +\cdots  &=(...)
 \\
\qquad t&=t_2:  &\quad (...)\re\, a_{0}(s) + (...)\re\, a_{2}(s) +  (...)\re\, a_{4}(s)  +\cdots  &=(...)
  \\
\qquad  &~~ \vdots & & \vdots~~~~~~~~~~~~~~~~~~~~~~& 
\end{aligned}
\ee
where the $\mu$ integrations on the right hand sides of the above equalities can be evaluated numerically for a given ansatz of $\im\, a_{\ell} (\mu)$. Specifically, we will sample $(\ell_{\rm max}/2 + 1)$ different points of $t$, evenly distributed\footnote{Note that this choice of sampling is made for pure convenience but does not affect the result.} in the region $[t_{\rm min}, t_{\rm max}]$, with $t_{\rm min}$ chosen to be close to $-4m^2$ and $t_{\rm max}$ close to $4m^2$. (These linear equations can be efficiently solved with the multi-precision linear algebra package MPLAPACK \cite{DBLP}.) To recapitulate, we only construct all possible $\im\, a_{\ell}(s)$'s, while $\re\, a_{\ell}(s)$'s are then obtained via analyticity.

Before specifying the exact unitarity conditions to be imposed, let us comment on the dependence of the spin truncation order $\ell_{\rm max}$ on $s$. As mentioned in Appendix~\ref{sec:largel}, the contributions of high-spin partial waves become more significant as $s$ increases. Thus, for a fixed $\ell_{\rm max}$, the truncated partial wave expansion is only reliable up to some $s_{\rm max}^{(\ell)}$. This may be understood as follows: For a massive scalar, its interactions are short-ranged, which means that if the impact parameter becomes large $b\sim \ell/\sqrt{s}\gtrsim m^{-1}$, the interactions become negligible. Fortunately, due to the $\mu$ suppression in the denominators, the integration over some scale $\mu > s_{\rm max}^{\rm int}$ is negligible. So for an accurate bootstrap, we need $s_{\rm max}^{(\ell)}> s^{\rm int}_{\rm max}$. Numerically, however, we find that we can only sample $s$ up to an $s_{\rm max}$ that is much smaller than $s_{\rm max}^{\rm int}$.

Therefore, for a given $\ell_{\rm max}$, we extract the real parts of the partial wave amplitudes via the dispersion relation up to $s_{\rm max}$, below which we can impose full partial wave unitarity. Numerically, we discretize the interval $[s_{\rm min}, s_{\rm max}]$, with $s_{\rm min}$ close to $4m^2$, and impose the following unitarity conditions
\begin{equation}
\label{SUnitarityConds}
    \begin{pmatrix}
        1 -\frac{1}{2}S_{m} \im\, a_{\ell}(s),&S_{m}^{1/2}\re\, a_{\ell}(s)\\
       S_{m}^{1/2}\re\, a_{\ell}(s), &2\im\, a_{\ell}(s)
    \end{pmatrix}
    \succeq 0 \,\text{, ~~~for $0\leq {\rm even}~ \ell \leq \ell_{\rm max}$,~ $s_{\rm min}\leq s \leq s_{\rm max}$\,,}
\end{equation}
where $S_{m}=((s-4m^2)/{s})^{1/2}$. From Eqs.~\eqref{imtoreal} and \eqref{ansatz_im}, we see that the partial wave amplitudes are linear functions of $\mathcal{M}(s_0, t_0)$ and $c_{\ell, k}$. These unitarity constraints are now standard linear matrix inequalities that can be implemented in a semi-definite program (SDP) problem, solvable by a package such as {\tt SDPB}. For $s>s_{\rm max}$, we can impose linear unitarity conditions for $\im\, a_\ell(s)$,
\be
0\leq S_m \im\, a_\ell(s)\leq 2\,,
\ee
which aids in improving convergence. It is worth pointing out that although full unitarity conditions are only imposed up to $s_{\rm max}$, to some extent, they impose constraints along the whole $s$ channel cut, because the dispersion relation already includes information along the cut.

The idea of parameterizing the amplitude via the dispersion relation has been studied in previous work. In Ref.~\cite{Bhat:2023puy}, a distinct parameterization scheme is explored based on the crossing-symmetric dispersion relation, the Roy equations and resonance structures above the threshold, implemented with a different numerical scheme that requires 2D integrations. As in the dual approach, the use of Roy's equations restricts the accessible range of $s$, whereas directly solving the (fixed $t$) dispersion relation for various values of $t$ circumvents this limitation.

\subsection{Trivilization of null constraints}\label{sec:null}

When extracting positivity bounds on EFT coefficients using the positivity parts of partial wave unitarity, $\im\, a_\ell(\mu)>0$, the null constraints from $st$ crossing symmetry can be used to obtain two-sided bounds for the ratios of EFT coefficients \cite{Tolley:2020gtv}. To get better bounds, one can additionally impose $tu$ crossing symmetry, which requires the odd spin partial waves to vanish. These two types of null constraints are independently effective in constraining the imaginary parts of the partial wave amplitudes, because the previous positivity bound formulation does not fully utilize analyticity and unitarity in the dispersion relation, particularly the information contained in $b_0(t)$.

In our current formulation, analyticity and unitarity are sufficiently utilized, this explains why we only need to impose either $tu$ or $st$ crossing symmetry, since the dispersion relation is already $su$-crossing symmetric. This can be checked by explicit numerical computations. If only a small number of analyticity and partial wave unitarity conditions are used, adding both types of null constraints may accelerate the numerical convergence, but the advantage disappears upon raising the numerical truncation order.

It is simplest to use $tu$ crossing symmetry, which simply requires that all odd-spin components in the partial wave expansion vanish in the physical region $s > 4m^2$. This is essentially because an analytic function is determined by its boundary values. Since the fixed-$t$ amplitude is an analytic function of $s$ whose boundary is two branch cuts linked by crossing symmetry, we only need to impose $tu$ crossing symmetry on the $s$-channel cut
\begin{equation}
    \mc{M}(s,t) = \mc{M}(s,4m^2-s-t) \text{, ~~~for $s> 4m^2$}\,.
\end{equation}
Once this condition is satisfied, the amplitude reconstructed using the dispersion relation will satisfy $tu$-crossing symmetry, and consequently $st$-crossing symmetry, on the entire complex $s$-plane. In our method, we construct all possible amplitudes by directly parametrizing the partial waves---the imaginary parts are parametrized and the real parts are inferred from the imaginary parts by solving algebraic equations. Thus, we can simply set all the odd-spin partial wave amplitudes to zero, without actually imposing them as constraints in the numerical evaluations. In other words, the null constraints are trivialized in this formulation.

\subsection{Numerical parameters}

In this subsection, we specify the choices of numerical parameters in the SDP implementation. As mentioned earlier, the dispersion relation can be imposed as constraints at discrete points of $s$ and $t$ in the numerical calculations. We will evenly draw $(\ell_{\rm max}/2 + 1)$ points of $t$ within the region $[t_{\rm min}, t_{\rm max}]$, with $t_{\rm min} = -(4 - 10^{-3})m^2$ and $t_{\rm max} = (4 - 10^{-3})m^2$, except in Section \ref{sec:glueball}, where the bound state poles will reduce the convergence region of the partial wave expansion and we will use a narrow $[t_{\rm min}, t_{\rm max}]$. Partial wave unitarity conditions are imposed at $N_s$ points within the region $s_{\rm min}\leq s\leq s_{\rm max}$, according to
\be\label{eq:disc}
s_k=s_{\rm min}+(s_{\rm max} -s_{\rm min}) \(\f{k}{N_s-1}\)^p,~~~k=0,1,...,N_s-1 \,,
\ee
with $s_{\rm min}$ chosen to be $ (4 + 10^{-4})m^2$ and $s_{\rm max}$ depending on $\ell_{\rm max}$. $p$ is chosen to be 1, 2 or 3, and we find that the numerical results for the $p=2$ or $p=3$ case converge with a smaller $N_s$, comparing to the even sampling case $p=1$. This reflects the fact that unitarity conditions in the lower energy region (smaller $s$ region) are more constraining for the leading couplings, as one may expect. Generally, a larger $\ell_{\rm max}$ allows for a greater $s_{\rm max}$, as discussed in Section \ref{sec:method}. Numerically, we may also need to increase $k_{\rm max}$ to reach convergence once a greater $s_{\rm max}$ is chosen. Physical results are obtained by raising $s_{\rm max}$ until convergence is reached. As a larger $s_{\rm max}$ is chosen, we should also increase the number of discrete points $N_s$. As mentioned, to accelerate convergence, we shall also impose linear unitarity conditions $0\leq S_m \im\, a_\ell(s)\leq 2$ for $s> s_{\rm max}$, where these discrete points are chosen as \be
s_i = s_{\rm max}/(1 - i/N'_s),~~~~i = 1, 2, \dots, (N'_s - 1) \,.
\ee

As for the choice of the two expansion orders in the ansatz for $\im \mc{M}(\mu,t)$, namely $k_{\rm max}$ and $\ell_{\rm max}$, it is generally advisable to first ensure a sufficiently large $k_{\rm max}$. This can be easily determined through a convergence study with fixed $\ell_{\rm max}$, as raising $k_{\rm max}$ generally increases the accuracy. If $k_{\rm max}$ is chosen too small, convergence with respect to $\ell_{\rm max}$ may appear rather elusive. This is related to the fact that the number of partial waves used in the current setup is matched to the number of discrete points of $t$ used to solve for the real parts of the partial waves---the $\ell_{\rm max}$ truncation is linked to two approximations. In any case, once a sufficiently large $k_{\rm max}$ is established, it is usually relatively straightforward to achieve convergence in $\ell_{\rm max}$. For all plots presented in this paper, convergence is sufficiently good, and therefore no extrapolation is performed.
See Table~\ref{tab:numerical_params} for a summary of the numerical parameters employed in this work. More details of the numerical implementation are given in Appendices \ref{sec:cal_int} and \ref{sec:cov_test}. 

\begin{table}[ht]
\centering
\caption{List of numerical parameters used in the method.}
\begin{tabular}{l|l}
\hline
\hline
$\ell_{\rm max}$  & Truncation order of partial wave expansion in \eqref{ansatz_im} \\
$k_{\rm max}$     & Truncation order of Legendre polynomials for each partial wave in \eqref{ansatz_im} \\
$s_{\rm max}$     & Upper limit of $s$ used in dispersion relation \eqref{imtoreal} with full unitarity\\
$N_s$     & Number of discretized $s$ used in dispersion relation \eqref{imtoreal} with full unitarity\\
$N'_s$     & Number of discretized $s$ used in dispersion relation \eqref{imtoreal} with linear unitarity\\
\hline
\end{tabular}
\label{tab:numerical_params}
\end{table}

\section{Bounds on amplitude coefficients}
\label{sec:bound_expansion_coeff}

In this section, we will compute the bounds on some of the lowest order amplitude coefficients for a massive scalar without bound states below the threshold. Specifically, we will expand the amplitude around the crossing symmetric point $s = t = 4m^2/3$ and calculate the bounds on the following two (scaled) taylor coefficients
\begin{equation}
\begin{aligned}
\label{g0g2coefs}
    g_0 = \frac{1}{32\pi}\mc{M}\Big(\f{4m^2}3,\f{4m^2}3\Big)\,,~~~~~~~
    g_2 = \frac{m^4}{32\pi} \frac{\partial_s^2\mc{M}\big(s,\f{4m^2}3\big)\big|_{s\to 4m^2/3}}{4}\,.
\end{aligned}
\end{equation}
We will also discuss the Regge (large $|s|$) behaviors of the amplitude at some extremal points of the bounds, and compare them to the corresponding behaviors that can be extracted from the previous primal method. The Jin-Martin bound \cite{Froissart:1961ux, Martin:1962rt,Jin:1964zza} states that as $|s|$ goes to infinity the scattering amplitude is bounded by $|{\cal M}(s\to\infty,t)|< C|s|^{1+\eta(t)}$, $C$ being constant and $0<\eta(t)<1$, for $t<4m^2$. We will explore how large $\alpha(t)$ can be and its dependence on $t$.

\subsection{Bounds on $g_0$ and $g_2$}\label{sec:bound_g0g2}

Let us first calculate the 2D bound on the coefficients $g_0$ and $g_2$. In terms of the decision variables in the parametrized dispersion relation introduced in the last section, $g_0$ is simply $\mc{M}(s_0,t_0) / 32\pi$, while $g_2$, or other general $g_i$, is a linear combination of the coefficients $c_{\ell,k}$. (To obtain the linear combination of $c_{\ell,k}$ for $g_i$, we simply expand the parametrized dispersion relation around $(s_0, t_0)$, match the two sides of the dispersion relation at each order, and compute the $\mu$ integrations numerically. Note that the $\mu$ integrations here are different from those of Appendix \ref{sec:cal_int}, independent of the numerical setup for $s_{\rm max}$, $N_s$ and so on, and thus are easy to compute.) Therefore, the problem of finding the bounds on $g_0$ and $g_2$ becomes an SDP with the decision variables $\mc{M}(s_0, t_0)$ and $c_{\ell,k}$, with the linear matrix inequalities being the unitarity conditions discussed in the previous section.

\begin{figure}
    \centering
    \includegraphics[width=0.65\linewidth]{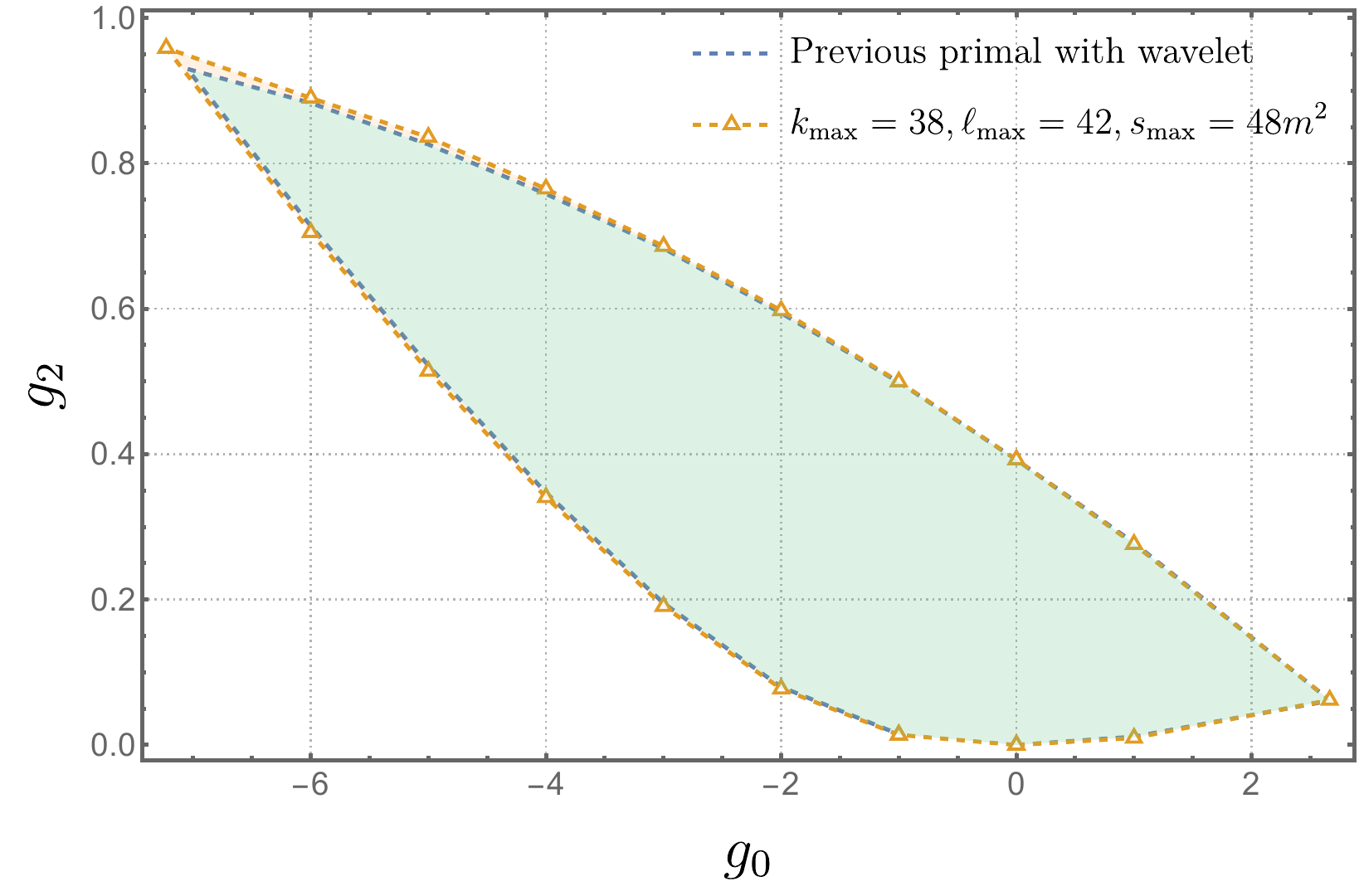}
    \caption{2D bounds on $g_0$ and $g_2$. The region with $k_{\rm max}=38$ has converged well, as evidenced by the convergence study in Figure \ref{fig:cov_g0min}. The previous primal bound is taken from Ref.~\cite{EliasMiro:2022xaa}, which uses an improved full amplitude ansatz with the wavelet technique (compared to that of Appendix \ref{sec:preprimal}).  
    }
    \label{fig:2d_g0g2}
\end{figure}

To find the boundary of the 2D bound, we first find a point $(g_0^{(0)}, g_2^{(0)})$ within the bounds. For example, we can choose $g_0^{(0)}=0$, compute the 1D bound on $g_2$ and then choose $g_2^{(0)}$ to be within the 1D bound. Then, the 2D bound can be obtained by setting 
\be
g_0 = g_0^{(0)} + R \cos{\theta}~~~{\rm and}~~~g_2 = g_2^{(0)} + R \sin{\theta} \,.
\ee
and maximizing $R$ for many different $\theta$'s. Since $g_0$ and $g_2$ are related by $(g_0-g_0^{(0)}) \sin{\theta} = (g_2-g_2^{(0)}) \cos{\theta}$, we can add this equality as a constraint and simply maximize $(g_0 - g_0^{(0)}) / \cos{\theta}$ if $\cos{\theta} \neq 0$; otherwise, we maximize $(g_2 - g_2^{(0)}) / \sin{\theta}$. 
The optimized 2D bound is shown in Figure \ref{fig:2d_g0g2}. We see that the coefficient $g_2$ reaches its upper bound as $g_0$ reaches its lower bound and vice versa. This 2D bound agrees very well with the previous primal result, except near the left hand kink, where the numerical convergence is slow and a noticeable improvement is observed. By directly using the ansatz of Appendix \ref{sec:preprimal}, we find that the lower bound of the left hand kink is $g_{0,\rm min}\simeq 6.6$, in agreement with the bound in Ref.~\cite{Paulos:2017fhb}. An improved ansatz involving a wavelet technique is implemented in Ref.~\cite{EliasMiro:2022xaa}, where a stronger bound on $g_{0,\rm min}$ is obtained. Our result provides a further small improvement near the left hand kink, as shown in Figure \ref{fig:2d_g0g2}.

The numerical bound converges faster near the upper limit of $g_0$ because, as we will see in the next subsection, near this kink the amplitude approaches a constant as $|s|$ increases and the $\ell=0$ partial wave dominates the amplitude.  On the other hand, more numerical resources are required to achieve convergence for the lower limit of $g_0$. Specifically, near the lower limit of $g_0$, the upper curve converges more slowly than the lower one. Nevertheless, from the convergence study in Figure \ref{fig:cov_g0min}, we see that the $k_{\rm max}=38$ bound in Figure \ref{fig:2d_g0g2} has well converged. Near the lower limit of $g_0$, the amplitude diverges as $|s|$ goes to infinity, which is made possible by a number of partial waves being sufficiently activated as $|s|$ becomes large. In fact, we expect that $g_0$ is constrained to be positive if we assume that the amplitude decays to zero at complex infinity, in which case we can use a non-subtracted dispersion relation (see Section \ref{sec:2rsubg0g2}). In contrast, $g_2$ has a positive sum rule, so the positivity part of unitarity already guarantees its positivity.

\begin{figure}
    \centering
    \includegraphics[width=1\linewidth]{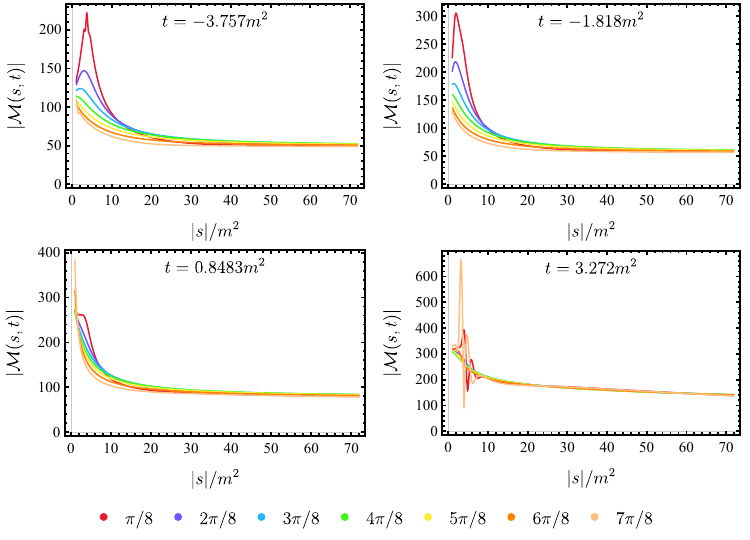}
    \caption{Large $|s|$ behaviors of the amplitude when $g_0$ reaches its upper bound. The numerical parameters are $\ell_{\rm max}=66, s_{\rm max} =72m^2, k_{\rm max} = 42$. The different colors denote different arguments $\thi$ on the complex $s$-plane: $s=|s|e^{i\thi}$.}
    \label{fig:g0_upper}
\end{figure}

\subsection{Regge behaviors} \label{sec:regge}

Our constructive method has the Jin-Martin bound built-in via use of the dispersion relation.  Now, let us examine the Regge behaviors at some extremal points of Figure \ref{fig:2d_g0g2}. At the boundary of the S-matrix bound, the coefficients $\mc{M}(s_0,t_0)$ and $c_{\ell,k}$ can be uniquely determined, so we can reconstruct the amplitude on the complex $s$-plane by the dispersion relation. However, due to the finite order truncation of the partial waves, we can only reliably reconstruct the amplitude up to $s \leq s_{\rm max}$ on the branch cut. This in turn means that the obtained amplitude is reliable within $|s| \lesssim s_{\rm max}$ on the complex plane. We shall survey the Regge behaviors of the amplitude along different directions in the complex $s$-plane.

First, we consider the kink where $g_0$ reaches its upper bound. In this case, it seems that the norm of the fixed-$t$ amplitude approaches a ``constant'' for large $|s|$, as shown in Figure \ref{fig:g0_upper}. (However, it is possible that the amplitude actually decreases slowly to zero as $s$ asymptotes to $\infty$, as the upper bound of $g_0$ can also be obtained from a modified primal bootstrap ansatz where the fixed-$t$ amplitude decays to zero at large $s$ \cite{EliasMiro:2022xaa}.) Although approaching the constant with different rates, the amplitude settles at almost the same constant for different arguments of complex $s$, but this constant varies for different $t$. We find that for $t>0$ this constant increases monotonically with $t$.

\begin{figure}
    \centering
    \includegraphics[width=0.9\linewidth]{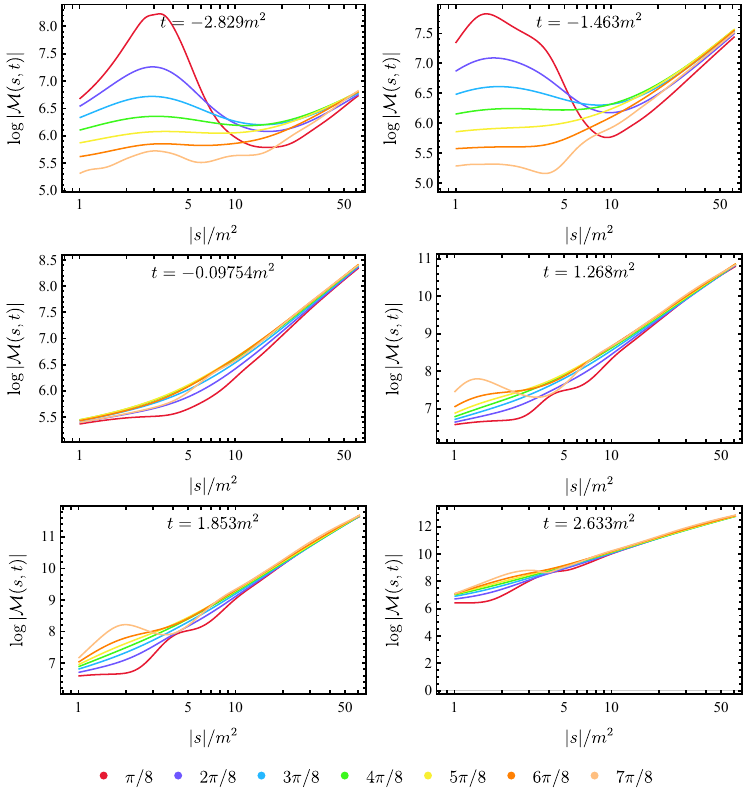}
    \caption{Large $|s|$ behaviors of the amplitude when $g_0$ reaches its lower bound. Here we choose $\ell_{\rm max}=82, s_{\rm max} =72m^2 , k_{\rm max} = 42$. The logarithm is with base $e$.}
    \label{fig:g0_lower}
\end{figure}

In contrast, the amplitude of the kink at the lower bound of $g_0$ diverges for large $|s|$, as shown in Figure \ref{fig:g0_lower}. Again, although the amplitude varies in its rates to approach the divergence along different directions in the $s$-plane, the final divergence rate at large $|s|$ is almost the same for all directions. As $t$ increases, this universal divergent behavior is reached at smaller values of $|s|$.

\begin{figure}
 	\begin{minipage}[b]{0.527\columnwidth}
 		\centering
 		\subfloat[$g_0$ at its lower-bound kink]{\includegraphics[width=1\linewidth]{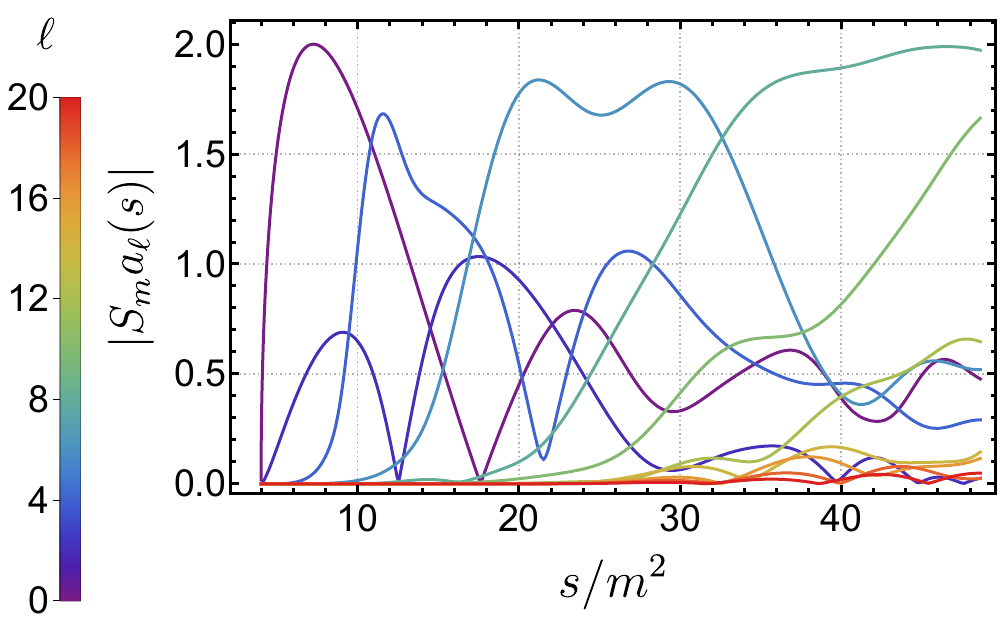}
 		
 	}
 		
 	\end{minipage}
 	\begin{minipage}[b]{0.473\columnwidth}
 		\centering
 		\subfloat[$g_0$ at its upper-bound kink]{\includegraphics[width=1\linewidth]{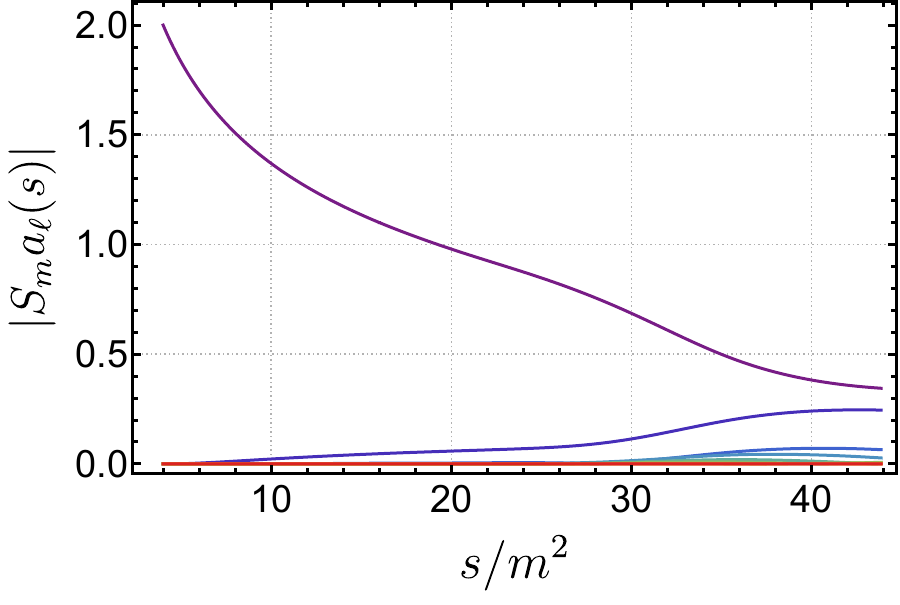}
 				
 		}
 	\end{minipage}
 	\caption{Partial wave amplitudes when $g_0$ reaches its lower and upper bound. The numerical parameters are $k_{\rm max} = 42 ,s_{\rm max} = 72m^2, \ell_{\rm max} =82$ (lower-bound kink) and $\ell_{\rm max} =66$ (upper-bound kink). Different colors represent different spins of the partial waves. As $|s|$ increases, the high-spin partial waves become more important if the (whole) amplitude diverges as $|s|\to\infty$. }
 	\label{fig:g0_partialwaves}
\end{figure}

Since each partial wave amplitude is bounded by unitarity, even at large $s$, achieving the divergent behavior of the full amplitude requires high-spin partial waves to become significant at large $s$. Figure \ref{fig:g0_lower} confirms that this is indeed the case when $g_0$ reaches its lower bound. On the other hand, if the amplitude remains bounded at large $s$, the high-spin partial waves are strongly suppressed. In Figure \ref{fig:g0_partialwaves}, we see that, for the amplitude at the upper bound of $g_0$, the first few partial waves dominate even for large $s$. By comparison, for the amplitude at the lower bound of $g_0$, the contributions of the high-spin partial waves become more significant as $s$ increases.

\begin{figure}
        \centering
        \includegraphics[width=0.89\linewidth]{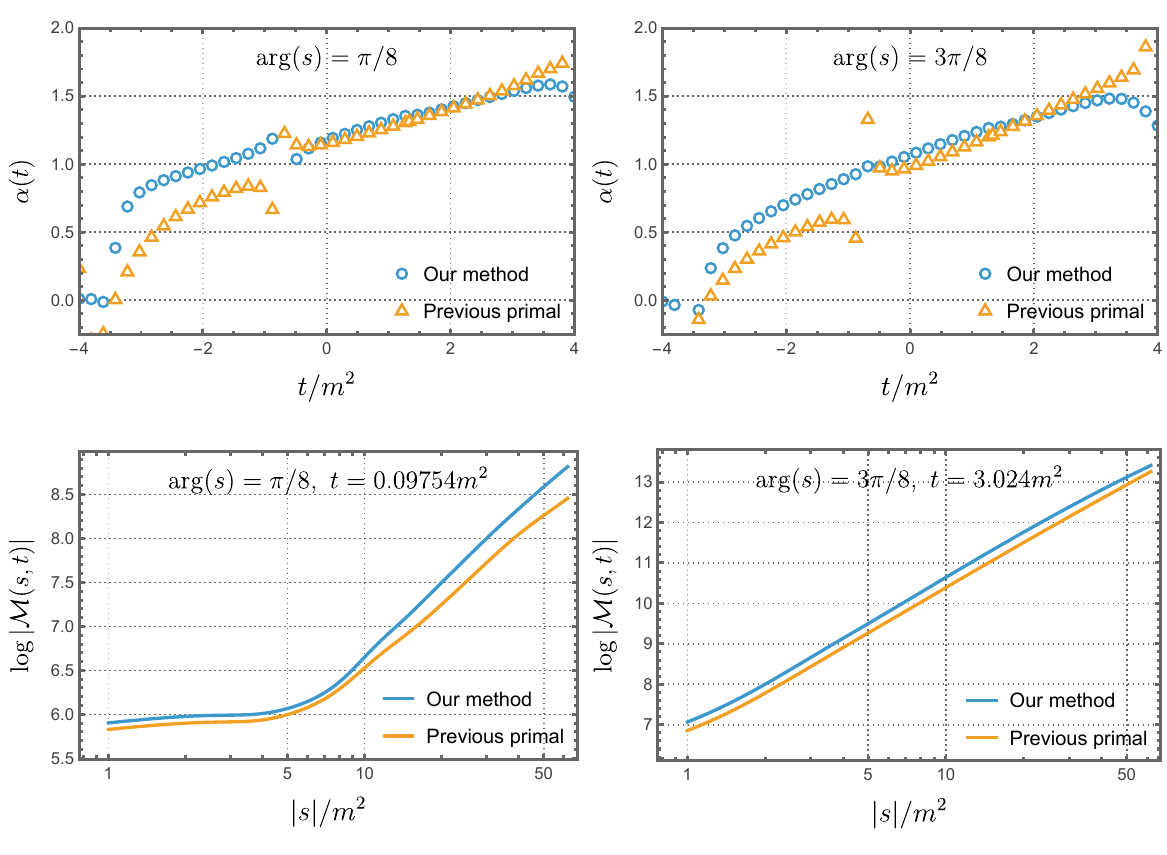} 		
 	\caption{(\textit{Upper panel}) $\alpha(t)$ fitted for the amplitude at the kink where $g_0$ reaches its lower bound. (\textit{Lower panel}) Reconstructed amplitudes at the kink for two choices of $t$. ${\rm arg}(s)$ is the complex angle of $s$.
    The ``previous primal" refers to the method of \cite{Paulos:2017fhb}. For our method, we take $\ell_{\rm max} =82, ~k_{\rm max} = 42, ~s_{\rm max} = 72 m^2$, while for the primal method of \cite{Paulos:2017fhb} we take $N_{\rm max} = 20$ (See Appendix \ref{sec:preprimal}).
    }
    \label{fig:cmp-std-1}
\end{figure}

To further probe the asymptotic characteristics of the kink amplitude at the lower-bound of $g_0$, we examine whether the amplitude can be fitted by the following form of the amplitude
\begin{equation}
    \mc{M}(s,t)\sim  s^{\alpha(t)}~~~{\rm for ~large~}s \,.
\end{equation}
Note that the amplitudes re-constructed from our numerical solutions at finite $|s|$ depend weakly on the argument of $s$. For concreteness, let us choose two representative arguments and the range $28m^2<s<36m^2$ for the fitting. For a comparison, we also numerically construct this kink amplitude using the previous primal method of \cite{Paulos:2017fhb}.  (The large $s$ behaviors of cross sections have recently studied with the previous primal method in Ref.~\cite{Correia:2025uvc}.) We find that the results obtained from our method closely resemble those from the previous primal approach. However, in general, the amplitude produced by our method tends to be slightly larger. In Figure \ref{fig:cmp-std-1}, we see that both methods exhibit sizable numerical errors near $t= 4m^2$, while the method of \cite{Paulos:2017fhb} seems to encounter notable numerical inaccuracies near $t=m^2$. Indeed, the fitted $\alpha(t)$ from the previous primal method tends to slightly overshoots the Jin-Martin bound near $t= 4m^2$, whereas in our approach it remains consistently within the bound, as the bound is hardwired via the dispersion relation. To reduce the errors near $t= 4m^2$, a greater $s_{\rm max}$ is required. However, imposing full unitarity with very large $s_{\rm max}$ is challenging with our current numerical implementation. Nevertheless, linear unitarity can be easily imposed at large $s$, which can lead to noticeable improvement. A clear trend is that the two methods show better agreement for $t>0$. Also, we see that in both methods the re-constructed amplitude diverges faster than ${\cal M}\sim s$ when $t\gtrsim -m^2$, reaching its fastest growth near $t=4m^2$, where it behaves approximately as ${\cal M}\sim s^{1.6}$ for the kink at the lower bound of $g_0$. It is worth noting that for $t < 0$, the amplitude tends to Reggeize at larger $s$, especially as $t$ becomes more negative (see Figure~\ref{fig:g0_lower}), making the fitted $\alpha(t)$ less accurate due to the limitation of finite $s_{\rm max}$. It would be interesting to pursue further investigations to better ascertain whether $\alpha(t)$ does indeed remain quasi-linear in this region which would be an interesting insight in its own right.

\begin{figure}
\begin{center}

\begin{minipage}[b]{1\textwidth}
    \centering
  \includegraphics[width=\linewidth]{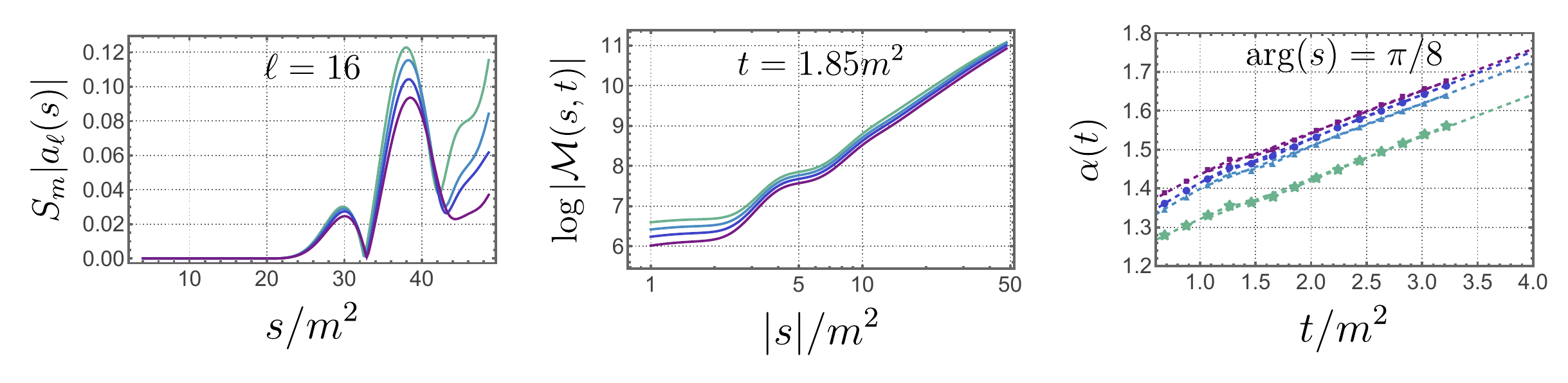}
\end{minipage}

\begin{minipage}[b]{0.6\textwidth}
  \centering
  \includegraphics[width=\linewidth]{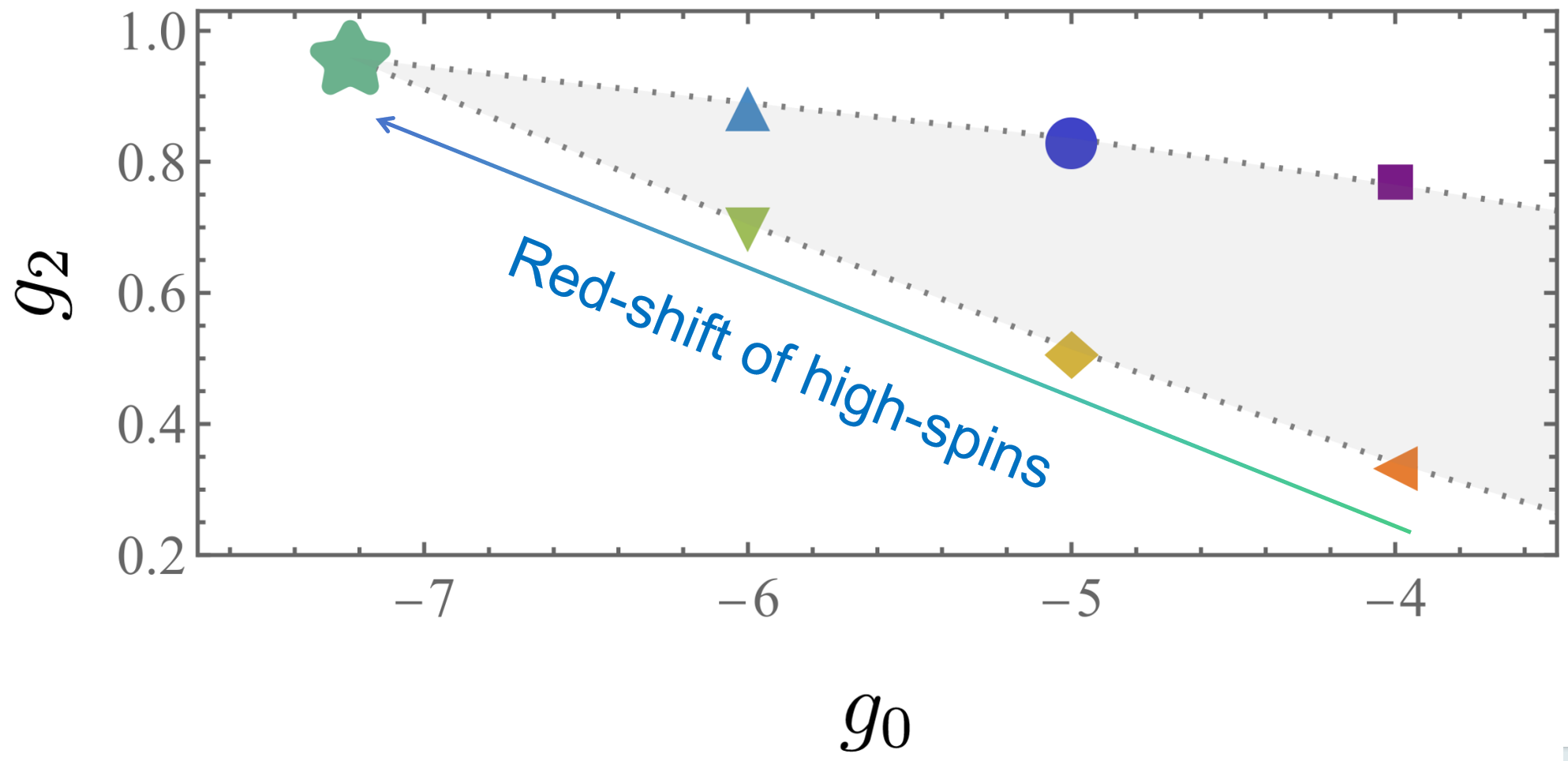} 
\end{minipage}

\begin{minipage}[b]{1\textwidth}
    \centering
  \includegraphics[width=\linewidth]{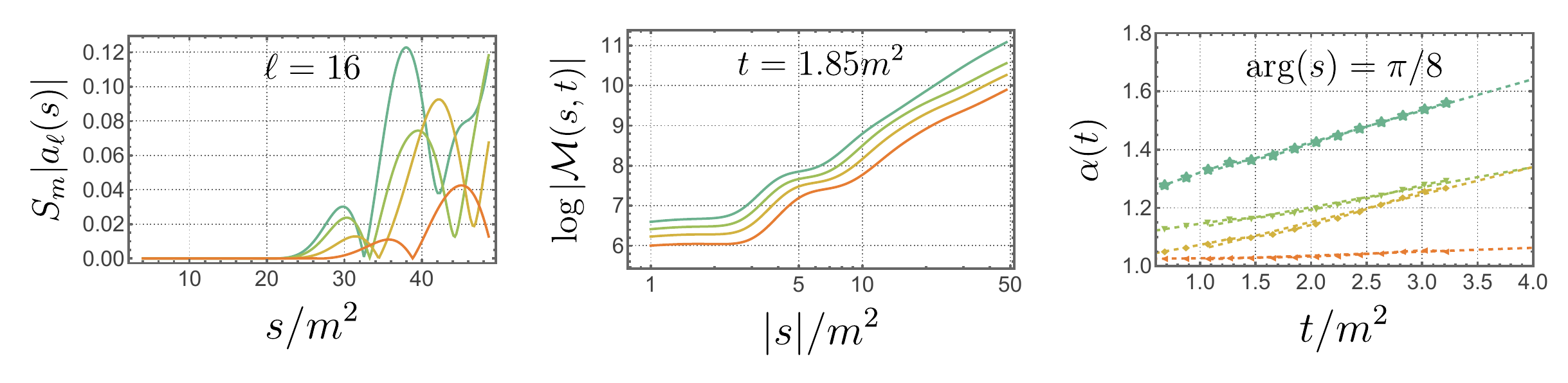}
\end{minipage}
\end{center}
\caption{Regge behaviors away from the left kink. A few points (with different colors) along the upper ({\it upper panel}) and lower ({\it lower panel}) boundary are chosen. The behavior of an exemplary high-spin partial wave ($\ell=16$ as an example) is also shown. The large $|s|$ behavior of the amplitudes is plotted with $t = 1.85m^2$ and ${\rm arg}(s) = \pi/8$. $\alpha(t)$ is fitted with data  within $28m^2<s<36m^2$. The numerical parameters are: $\ell_{\rm max} = 82$, $k_{\rm max} = 42$, and $s_{\rm max} = 72m^2$. In the ${\rm Re}\,\alpha$-$t$ plots, the dashed segments without data points are extrapolated from $1.07m^2\leq t \leq3.2m^2$ using linear fits.}
\label{fig:away_g0min}

\end{figure}

Let us comment on why the lower boundary curve in the $(g_0, g_2)$ parameter space converges more rapidly than the upper one. In Figure~\ref{fig:away_g0min}, we examine the amplitudes at various points along both the upper and lower boundary curves as we move away from the left hand kink. All amplitudes shown in this figure exhibit divergent behavior at large $|s|$, indicating that high-spin partial waves eventually become significant. However, as we move along the lower curve away from the kink, the onset of high-spin contributions shifts to increasingly larger values of $|s|$. To illustrate this, in Figure~\ref{fig:away_g0min}, we plot the spin-16 partial waves (as an example), the fixed-$t$ amplitudes and the fitted growth rates $\alpha(t)$ (extracted from the range $28m^2<s<36m^2$). In contrast, along the upper curve, this shift is much less pronounced, which explains the fact that it is numerically harder to achieve convergence along the upper curve. We see that the fastest Regge growth of the amplitude can reach at least $s^{1.7\sim 1.8}$ near $t=4m^2$ as we move away from the left kink along the upper curve.

These observed growth rates, quantified by the exponent $\alpha(t)$, can be interpreted within the framework of Regge theory. In this formalism, a trajectory of resonance states in the $t$-channel corresponds to a pole in the complex angular momentum plane at $\ell = \alpha(t)$. This implies that for a spin-$\ell$ resonance at a mass $m_\ell$ satisfying 
\be
\alpha(m_\ell^2) = \ell,
\ee
the associated contribution to the $s$-channel amplitude asymptotically scales as $s^{\alpha(t)}$. The Regge trajectory that governs the leading high-energy behavior of the amplitude is sometimes known as the Pomeron, originally proposed in the context of QCD. Consequently, the function $\alpha(t)$ that we have fitted for $t < 4m^2$ characterizes this trajectory. If the Pomeron trajectory is approximately linear, this empirical result can be extrapolated to predict the spectrum of higher-spin resonances in the $t$-channel. Owing to the crossing symmetry of the amplitude, these predictions are equally valid for resonance states in the direct $s$-channel.

\begin{figure}
    %\centering
    \includegraphics[width=1\linewidth]{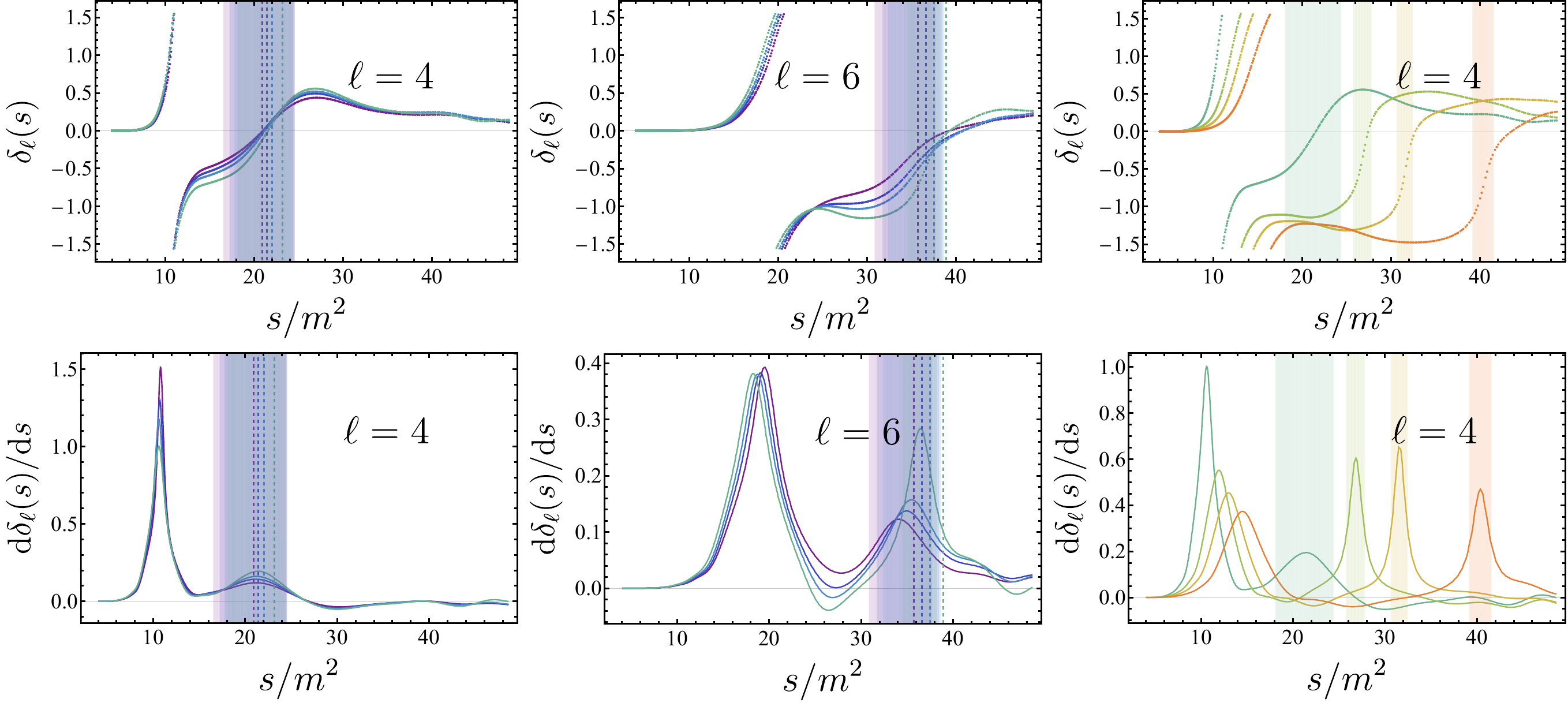}
    \caption{Phase shifts $\delta_\ell(s)$ and their derivatives for partial waves along the boundary curve, with colors corresponding to distinct points on the curve in Figure \ref{fig:away_g0min}. Shaded regions indicate the locations of Pomeron-like resonances identified from the phase shifts, with their widths given by the half maximum of the peak in ${\rm d} \delta_\ell(s)/ {\rm d}s$. Dashed lines in the left two columns represent resonance predictions from the fitted Regge trajectory $\alpha(t)$ over the interval $1.07m^2 < t < 3.2m^2$.}
    \label{fig:Pomeron_upper}
\end{figure}

Indeed, this picture is confirmed by our analysis of the optimal amplitudes on the boundary. As we proceed along the upper curve of the boundary, we find that the growth rate $\alpha(t)$ is approximately linear for $0 < t < 4m^2$. We have fitted $\alpha(t)$ in the region $1.07m^2 < t < 3.2m^2$ and used the resulting trajectory to predict the locations of the resonance states, which are shown as the dashed lines in the left panel of Figure~\ref{fig:Pomeron_upper}. (We average the fitted $\alpha(t)$ over seven directions in the complex $s$-plane ($\arg(s) = k\pi/8$, $k=1,\dots,7$) to reduce artifacts from the finite-$s$ analysis.) The shaded regions in the same figure indicate the resonance states identified from the fast variation of the partial-wave phase shift $\delta_\ell(s)$:
\be
\delta_\ell=\f12 {\rm arg}(1+iS_ma_{\ell}(s))\,.
\ee
The consistency between these two determinations supports the conclusion that the resonance states corresponding to the second peaks in the derivative $\mathrm{d}\delta_\ell(s)/\mathrm{d}s$ form a Pomeron trajectory. The trajectory at the left kink, which is responsible for the dips in the total cross-section, was also recently identified as Pomeron-like in \cite{Correia:2025uvc}, based on a different method that first fitted the resonance states and then computed the resulting amplitude growth rate at $t=0$.

In contrast, the predictive accuracy of this linear trajectory model diminishes along the lower curve of the boundary. This discrepancy could be a numerical artifact of the finite-$s$ fitting procedure, or it may indicate a genuine deviation from linearity of the Regge trajectories for amplitudes on the lower boundary. Nevertheless, we conjecture that these trajectories still lead to the leading asymptotical divergence in $s$, since the systematic shift of resonance states to higher energies when moving away from the kink (see the right column of Figure~\ref{fig:Pomeron_upper}) is consistent with the observed decrease in the growth rate of $\alpha(t)$ with increasing $t$.

\section{Massive scalar with cubic couplings}
\label{sec:spin-0}

In the previous sections, we have assumed for simplicity that the massive scalar theory does not contain cubic self-couplings. In this section, we will study bounds on a massive scalar with cubic self-couplings, and we will see that our method discussed above can be easily generalized.

Cubic couplings lead to the amplitude having poles at $s,t,u=m^2$. For fixed $t$, there are the $s$ channel pole at $s=m^2$ and the $u$ channel pole at $s = 3m^2-t$. For scalar scattering, the residues of these poles are constants, which are equal for different channels due to crossing symmetry and are positive due to unitarity, so we can denote them as $g^2$. The existence of these poles in principle will shrink the valid $t$ region of the dispersion relation. However, for a scalar amplitude, it is easy to parameterize the fully crossing symmetric pole part and separate it from the rest of the amplitude, which is again a function with only branch cuts. 

To see this, note that the fully crossing symmetric ansatz for the pole term is 
\begin{equation}
\label{Mpoledefg}
    \mc{M}_{\rm pole}(s,t):=g^2\left(\frac{1}{m^2-s} + \frac{1}{m^2-t} + \frac{1}{m^2-u} \right)\,,
\end{equation}
and we can define the pole-subtracted amplitude as
\begin{equation}
\label{McutMMpole}
    \mc{M}_{\rm cut}(s,t) = \mc{M}(s,t) - \mc{M}_{\rm pole}(s,t)\,.
\end{equation}
For fixed $s\geq 4m^2$, we can derive the imaginary parts of the partial waves for $\mc{M}_{\rm cut}(s,t)$ by multiplying with Legendre polynomials and then integrating over $t$ from $4m^2-s$ to 0:
\bal
    \im a^{\rm cut}_\ell(s) &= \frac{1}{16\pi (s-4m^2)} \int_{4m^2-s}^0 \d t P_\ell\left(1+\frac{2t}{s-4m^2}\right) \im \mc{M}_{\rm cut}(s,t) \\&= \frac{1}{16\pi (s-4m^2)} \int_{4m^2-s}^0 \d t P_\ell\left(1+\frac{2t}{s-4m^2}\right) (\im \mc{M}(s,t) - \im \mc{M}_{\rm pole}(s,t)) 
    \\&= \frac{1}{16\pi (s-4m^2)} \int_{4m^2-s}^0 \d t P_\ell\left(1+\frac{2t}{s-4m^2}\right) \im \mc{M}(s,t)  ,~~~s\geq 4m^2 \,,
\eal
where in the second equality we have used \eref{McutMMpole}, and in the third equality we have used the fact that the imaginary part of $\mc{M}_{\rm pole}(s,t)$ vanishes when $4m^2-s \leq t\leq 0$. Therefore, $\mc{M}_{\rm cut}$ and $\mc{M}$ have the same imaginary parts in the partial wave expansion in the physical region. They differ in the real parts of the partial waves amplitudes, which leads to different convergence regions: The partial wave expansion for $\mc{M}_{\rm cut}(s,t)$ converges for $|t|<4m^2$ and $s\geq 4m^2$, while that of $\mc{M}(s,t)$ converges for $|t|<m^2$ and $s\geq 4m^2$. As $\mc{M}_{\rm cut}$  is polynomial bounded and has the same imaginary parts of the partial waves as $\mc{M}$, it is easy to see that $\mc{M}_{\rm cut}$ satisfies the Froissart bound.

It follows that $\mc{M}_{\rm cut}(s,t)$ obeys the same analytic properties, the same region of convergence for the partial wave expansion and one can use the same ansatz for its partial waves as for the scattering amplitude without poles. Denoting the partial wave amplitudes of $\mc{M}_{\rm cut}$ and $\mc{M}_{\rm pole}$ as $a^{\rm cut}_\ell$ and $a^{\rm pole}_\ell$ respectively, to perform the bootstrap, the only difference is that now the unitarity conditions become 
\begin{equation}
\begin{aligned}
    &\begin{pmatrix}
        1 -\frac{1}{2} S_m\im\, a^{\rm cut}_{\ell}(s),&S_m^{1/2}(\re\, a^{\rm cut}_{\ell}(s)+\re\, a^{\rm pole}_\ell(s))\\
        S_m^{1/2}(\re\, a^{\rm cut}_{\ell}(s)+\re\, a^{\rm pole}_\ell(s)), &2\im\, a^{\rm cut}_{\ell}(s)
    \end{pmatrix}
    \succeq 0\\&~~~~~~~~~~~~~~~~~~~~~~~~~~~~~~~~~~ \text{for even $0\leq \ell \leq \ell_{\rm max}$, $s_{\rm min}\leq s \leq s_{\rm max}$\,,}
\end{aligned}
\end{equation}
because these conditions are imposed above the threshold, where only the real parts of the partial wave amplitudes are affected by the pole contributions. Here, $\mc{M}_{\rm pole}(s,t)$ is known up to the constant $g^2$, so $\re\, a^{\rm pole}_{\ell}(s)$ can be easily computed. In the numerical optimization, we simply need to add $g^2$ as an extra non-negative decision variable.

\begin{figure}
\includegraphics[width=0.48\linewidth]{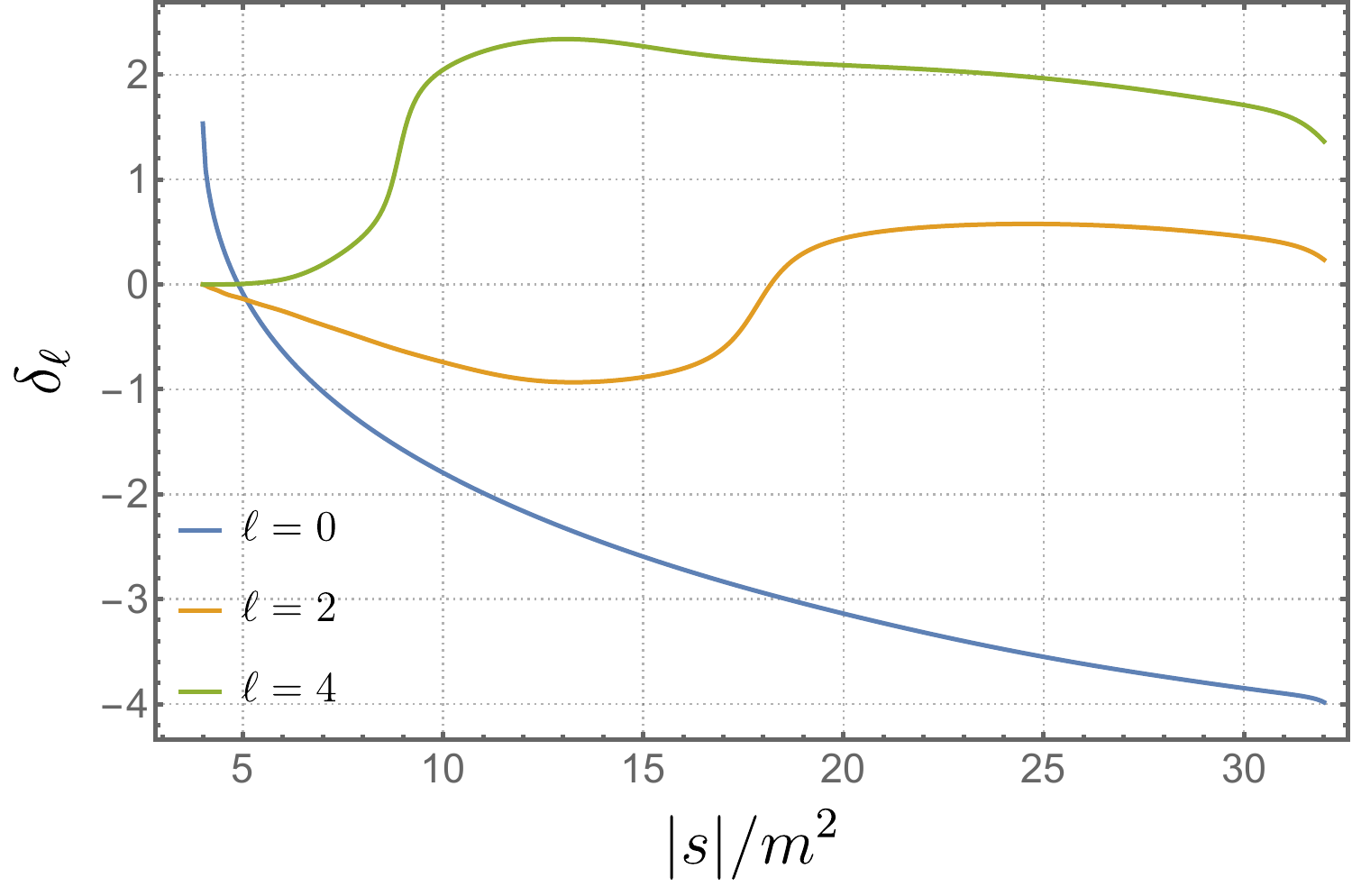}
~~~~
\includegraphics[width=0.48\linewidth]{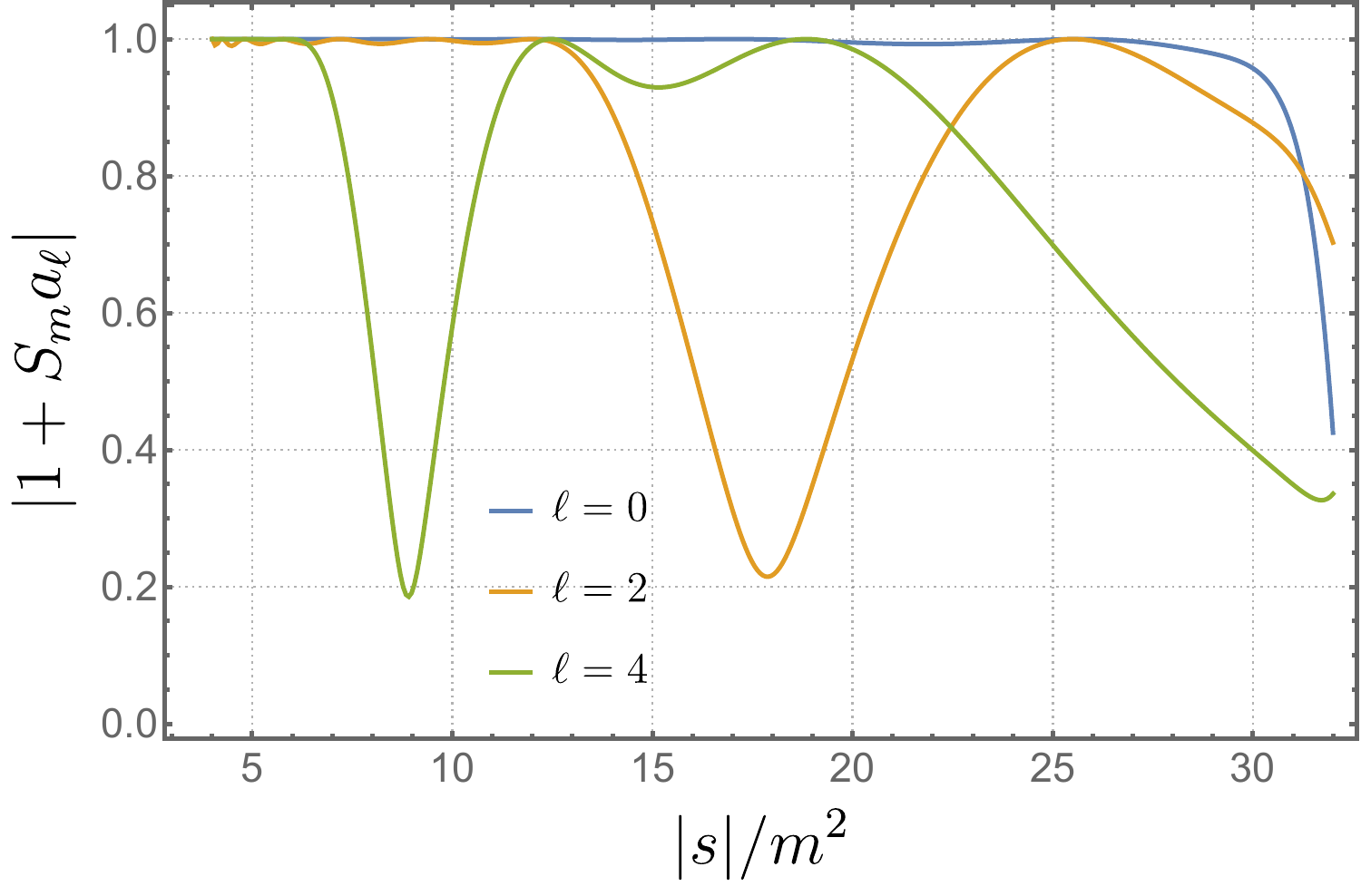}
 	\caption{Phase shifts ({\it left plot}) and absolute values of the S-matrix ({\it right plot}) of the partial waves when the coupling $|g|$ (see \eref{Mpoledefg}) reaches its upper bound.} 
 	\label{fig:spin_0_bounded_ps}
\end{figure}

Running the numerical optimization, we find that the bounds on the coupling constant $g$ are
\begin{equation}
    0\leq |g| < 48.8 m \,.
\end{equation}
This upper bound is consistent with the result from the primal method that constructs the full amplitude, which gives a bound: $0\leq |g|\lesssim 48m$ \cite{Paulos:2017fhb}. The phase shifts of the spin 0, 2 and 4 partial waves
are shown in the left panel of Figure \ref{fig:spin_0_bounded_ps}. In plotting the phase shifts, we have made use of the fact that their periods are $\pi$. We find that the phase shift of the spin-0 partial wave goes to $\pi/2$ at $s\to4m^2$, as a result of the formation of a threshold singularity. This singularity is captured in our $a_\ell(s)$ ansatz by the $\sqrt{s/(s-4m^2)}=1/S_m$ factor, which in turn cancels $S_m$ in the $\delta_\ell$ expression. Note that only the spin-0 partial wave is allowed to be singular at the threshold by unitarity and the existence of a dispersion relation (see \eref{aellos4m2}). Therefore, phase shifts of spin-$\geq 2$ partial waves must vanish at the threshold, a feature that is also accurately captured in the left plot of Figure \ref{fig:spin_0_bounded_ps}. The phase shift plot also reveals at least two rapid increases--one in the $\ell=2$ wave and another in the $\ell=4$ wave--corresponding to two resonances. These resonances are also clearly visible in the right plot of Figure \ref{fig:spin_0_bounded_ps}. The dips observed in the $\ell=2$ and $\ell=4$ waves are caused by zeros in their respective partial waves, located off the real $s$-axis on the physical Riemann sheet---these resonances manifest as poles in an unphysical Riemann sheet.

\section{Constraining glueball scattering} \label{sec:glueball}

Now let us apply our method to constrain 2-to-2 glueball scattering. Glueballs have the quantum numbers of scalar particles (with mass $m$), but in the 2-to-2 amplitude there are multiple bound states below $4m^2$, which appear as poles of the amplitude on the real $s$ axis. While the spin-0 bound state poles can be easily subtracted with a triple crossing symmetric ansatz, as we did in Section \ref{sec:spin-0}, such a procedure does not apply to the spin-$\ell\geq 2$ poles\,\footnote{In a triple crossing symmetric amplitude, there are only bound states with even $\ell$.}. These poles reduce the convergent region of the analytic dispersion relation, and thus a smaller range $t_{\rm min}\leq t \leq t_{\rm max}$ can be used. Extra contributions from these poles will also appear in the dispersion relation and change the large $\ell$ behavior of the partial waves. In this section, we will generalize our method so as to bootstrap the bounds on the coupling constants of the non-scalar bound states. 

\subsection{Adding spin-$\ell\geq 2$ bound states}

As discussed in Section \ref{sec:spin-0}, a spin-0 bound state can be easily handled by subtracting the pole part from the dispersion relation using a triple crossing symmetric ansatz and slightly modifying the unitarity conditions. Let us assume that this has been dealt with and there further exist spin-$\ell\geq 2$ bound states with masses $m_{b}$, which appear as poles in the amplitude at $s,t,u=m^2_b$, with $b$ labeling the bound states.

Because of the poles on the complex $t$-plane, which are located at $t=m_b^2$ ($t$ channel) and $4m^2-m_b^2-s$ ($u$ channel), the partial wave expansion diverges when $t$ hits the first/lightest pole at $t_*$. Also, the $t$ region where the Jin-Martin bound holds is reduced to
\begin{equation}
    \lim_{|s|\to\infty}|\mc{M}(s,t)|<|s|^{1+\gi(t)}\text{,~~~ for $|t|<t_*$ and $0<\gi<1$} \,.
\end{equation}
Therefore, we must choose $t_{\rm min}>-t_*$ and $t_{\rm max}<t_*$, where $t_{\rm min}$ and $t_{\rm max}$ are defined in Section \ref{sec:method}. The residues of the $s$ and $u$ channel pole are given by
\begin{equation}
    -g_b^2 P_{\ell}\left(1+\frac{2t}{m_b^2-4m^2}\right)\,,
\end{equation}
where $g_b$'s are the coupling constants of the bound states. So, with these poles added, the dispersion relation for a real $t$ is modified to
\begin{equation}
\begin{aligned}
    \mc{M}(s,t)&=\mc{M}(s_0,t_0) + \sum_b g_b^2\left(P_{\ell}\left(1+\frac{2t}{m_b^2-4m^2}\right) K^{m_b^2,t_0}_{s,t}+P_{\ell}\left(1+\frac{2t_0}{m_b^2-4m^2}\right)K^{m_b^2,s_0}_{t,t_0}\right)  \\
    &~~~~ +\int_{4m^2}^{\infty} \frac{\d \mu}{\pi}\left(\im \mc{M}(\mu,t) K^{\mu,t_0}_{s,t}  + \im \mc{M}(\mu,t_0) K^{\mu,s_0}_{t,t_0} \right)\text{, ~~for $-t_*<t<t_*$}\,,
\end{aligned}    
\end{equation}
where $t_0$ must be chosen inside the region $(-t_*,t_*)$. In this section, we choose $(s_0,t_0)=(2m^2,0)$.

Furthermore, the existence of bound states alters the large $\ell$ behavior of the partial waves. To see this, let us employ the Froissart-Gribov formula (with poles), which, for $s>4m^2$ and sufficiently large $\ell$, is given by
\begin{equation}
\begin{aligned}
    a_{\ell}(s)& = \frac{1}{32\pi}\sum_{b} \left(Q^{(4)}_{\ell}(z_b) \underset{z\to z_b}{{\rm Res}} \mc{M}(s,t(z_b)) + Q^{(4)}_{\ell}(-z_b) \underset{z\to -z_b}{{\rm Res}} \mc{M}(s,t(-z_b))\right) \\
    &
    ~~~~ + \text{(branch-cut contributions)}\,,
\end{aligned}
\end{equation}
where $t(z_b)=m^2_b$ is the pole of a $t$-channel bound state, meaning that $z_b = 1+2m_b^2/(s-4m^2)$, and $t(-z_b) = 4m^2-m_b^2-s$ is the pole of a $u$-channel bound state. The contributions from the branch cuts give the same large $\ell$ factor as specified in Appendix \ref{sec:largel}. Using $su$ crossing symmetry and $Q^{(4)}_{\ell}(-z) = (-1)^{\ell+1}Q_{\ell}^{(4)}(z)$, we see that the odd partial wave amplitudes vanish, and 
\begin{equation}
    a_{\ell={\rm even}}(s) = \frac{1}{16\pi} \sum_{b} Q^{(4)}_{\ell}(z_b) \underset{z\to z_b}{{\rm Res}} \mc{M}(s,t(z_b))+ \text{(branch-cut contributions)} \,.
\end{equation}
For a large $\ell$, from \eref{largel_Q}, we see that $Q_{\ell}^{(4)}$ contains the following factor 
\begin{equation}
    \left(z_b+\sqrt{z_b^2-1}\right)^{-\ell} = \left(\frac{s-4m^2}{s-4m^2+2m_b^2 + 2m_b\sqrt{s-4m^2+m_b^2}} \right)^\ell \,.
\end{equation}
For spin-$\ell \geq 2$ bound states, we generally expect $\underset{z\to z_b}{{\rm Res}} \mc{M}(s,t(z_b))$ to be complex, and thus we must modify the ansatz for the imaginary parts of the partial wave amplitudes. Although these factors also decay exponentially for large $\ell$'s, the decay is slower compared to the factor in \eref{factor_cut}. So it is beneficial to add such a factor to the ansatz for the imaginary parts of the partial wave amplitudes, which now becomes\,\footnote{For the $\ell =0$ component, the two pre-factors are equal to 1, so we can simply set $c^{(b)}_{0,k}=0$ without loss of generality.}
\begin{equation}\label{ansatz_im_pole}
\begin{aligned}
    \im\, a_{\ell}(s) &= \sqrt{\frac{s}{s-4m^2}}\bigg\{\left( \frac{s-4m^2}{s+4m^2+4m\sqrt{s}}\right)^\ell \sum_{k=0}^{k_{\rm max}} c_{\ell,k} P_k(8m^2/s -1)\\
    &~~~~ +\sum_b \bigg(\frac{s-4m^2}{s-4m^2+2m_b^2 + 2m_b\sqrt{s-4m^2+m_b^2}} \bigg)^\ell \sum_{k=0}^{k^{(b)}_{\rm max}} c^{(b)}_{\ell,k} P_k(8m^2/s -1) \bigg\}\,.
\end{aligned}
\end{equation}
Similar to the case in Section \ref{sec:method}, this ansatz ensures the convergence of the dispersion relation and captures the large $\ell$ behavior of the amplitude, while for small $\ell$'s the differences are unimportant, as they can be compensated by the $k$ expansion. Again, in the numerical optimization, the $g_b^2$ couplings are treated as decision variables that are positive, as a result of unitarity.

\subsection{Bounds on glueball couplings}

\begin{figure}
    \centering
    \includegraphics[width=0.6\linewidth]{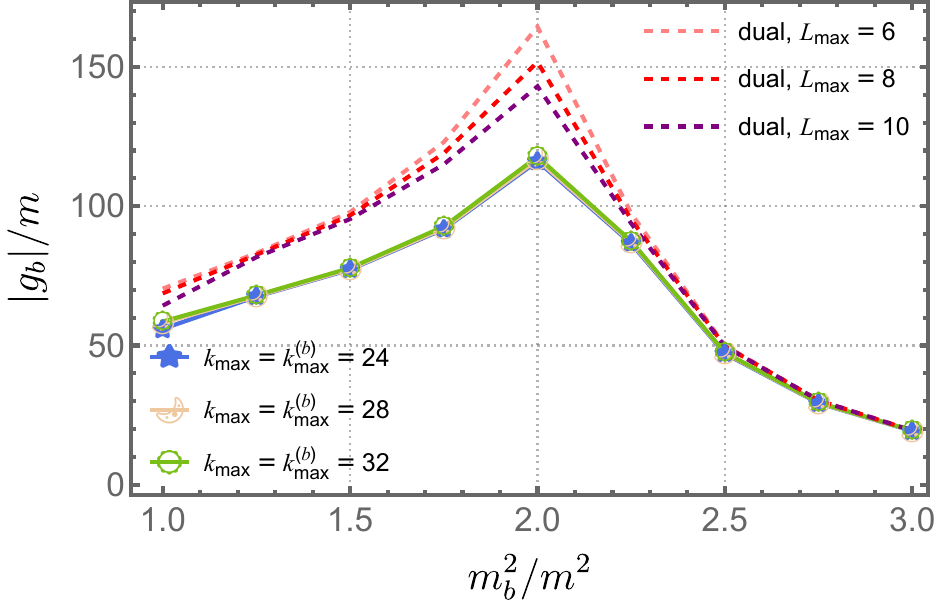}
    \caption{Upper bounds on the coupling constant $|g_b|$ for a spin-2 bound state, with $m_b$ being the bound state mass and $m$ being the glueball mass. Here we choose $\ell_{\rm max}=24$, $s_{\rm max} = 12m^2$ and $t_{\rm max} = -t_{\rm min} = 2m_b^2/3$, with different $k_{\rm max} = k_{\rm max}^{(b)}$. The dotted lines are extracted from the bounds obtained with the dual method of \cite{Guerrieri:2023qbg} for different $L_{\rm max}$ ($L_{\rm max}$ in the dual method is different from our $\ell_{\rm max}$). }
    \label{fig:spin2_pole}
\end{figure}

With the formalism we have provided so far, we can now apply it to compute the allowed region for the bound state couplings $g_b$. First, we apply our method to the case where there is only a single spin-2 bound state below the threshold. This acts as a sanity check, as the bound on this spin-2 pole coupling has been recently been computed by a dual bootstrap method \cite{Guerrieri:2023qbg}. The result is shown in Figure \ref{fig:spin2_pole}, where we observe excellent convergence in our method. We see that our bound is consistent with that of the dual bound \cite{Guerrieri:2023qbg}, but there is still a sizable duality gap for smaller values of $m_b$. Note that a similar duality gap also exists for the spin-0 bound state case between the previous primal method and the dual method \cite{Guerrieri:2023qbg}. As expected, the location of the peak at $m_b^2 = 2m^2$, corresponding to the case where the singularities of the $s$ and $t$ (or $u$) channel partial waves start to overlap, aligns well with that of the dual method. 

Next, let us use our method to constrain real world glueball couplings. Glueballs are massive, color-neutral, composite particles made of gluons in the low energy confinement phase of Yang-Mills theory, thanks to the nonlinear interactions between gluons. They can have different spins and masses. Following \cite{Guerrieri:2023qbg}, we shall use the following mass spectrum from lattice QCD simulations \cite{2007.06422, 2106.00364} as an input to bootstrap the bounds on the couplings:
\be
G(0): 1;  ~~~H(2): 1.437 \pm 0.006; ~~~G^*(0): 1.72\pm 0.01;~~~H^*(2): 1.99\pm 0.01 \,, \nonumber
\ee  
where the spins of the states are indicted in the brackets and the masses are normalized to the lightest (scalar) glueball $G$. While the non-perturbative lattice method can determine the mass spectrum relatively easily, extracting the couplings remains a challenging task there. We will disregard the state $H^*$, because its mass is near the 2-particle threshold and so the distinction between a bound state and a resonance becomes unclear. 

Consider identical scalar scattering between two $G$ glueballs. This process can probe the $G$ self coupling as well as its interaction with other glueballs, enabling the derivation of bounds on the residues at bound-state poles. Note that we have both spin-0 and spin-2 bound states below the threshold:
\begin{equation}
    \mc{M}_{GG\to GG}(s, t) \supset \sum_b\frac{g_b^2 P_{\ell_{b}}(1+2t/(s-4m^2))}{m_b^2 - s}\,,
\end{equation}
where the sum over $b$ includes the states $G$, $H$ and $G^*$ with $m_b$ and $\ell_{b}$ labeling their masses and spins respectively. For the scalar bound states $G$ and $G^*$, we shall use the method of Section \ref{sec:spin-0} to directly subtract the poles, while for the spin-2 bound state $H$ we use the parameterization of \eref{ansatz_im_pole} to account for it. The amplitude with the scalar poles subtracted has poles at $t = m_H^2$ and $t = 4m^2 -s- m_H^2$ on the $t$-plane, and is analytic at least within $|t| < m_H^2$. Therefore, we shall choose $t_{\rm min} = -m_G^2$ and $t_{\rm max} = m_G^2$ in our bootstrap implementation.

\begin{figure}
\includegraphics[width=0.48\linewidth]{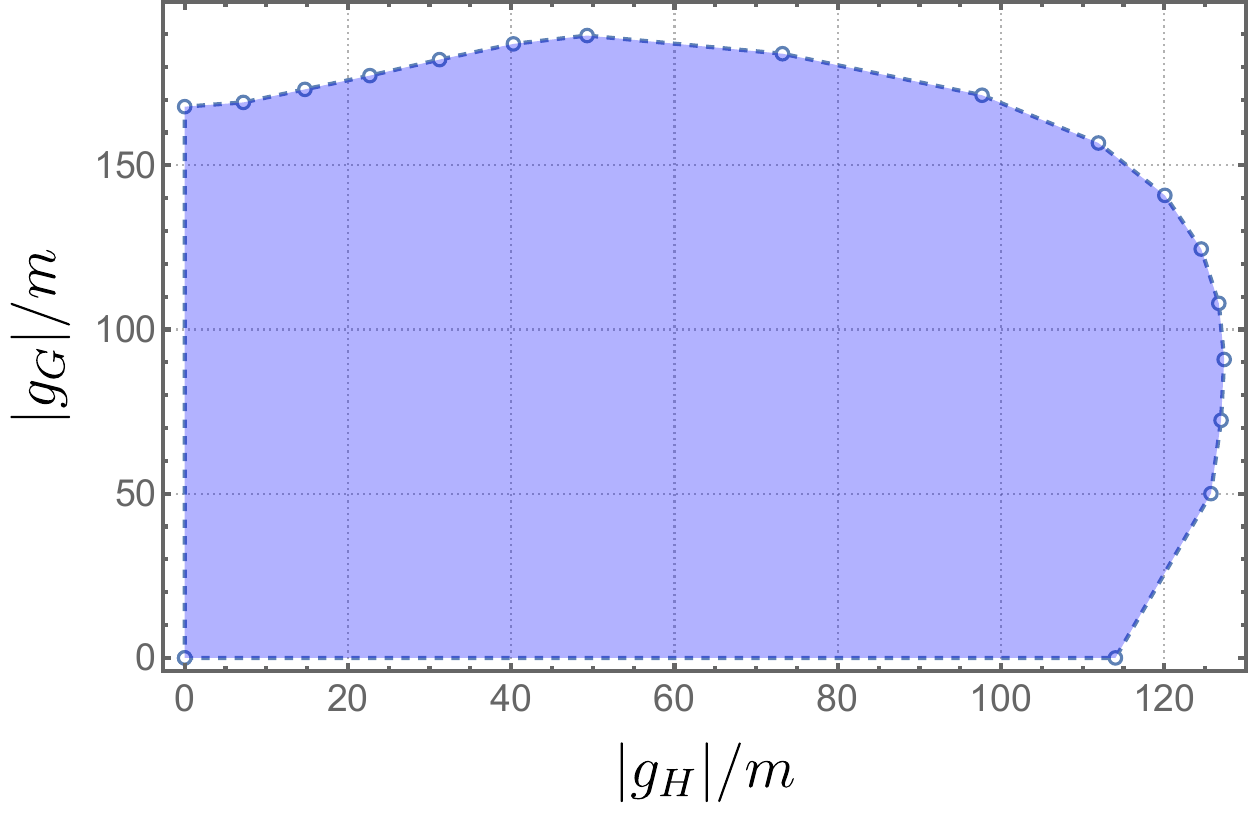}
~~~~
\includegraphics[width=0.48\linewidth]{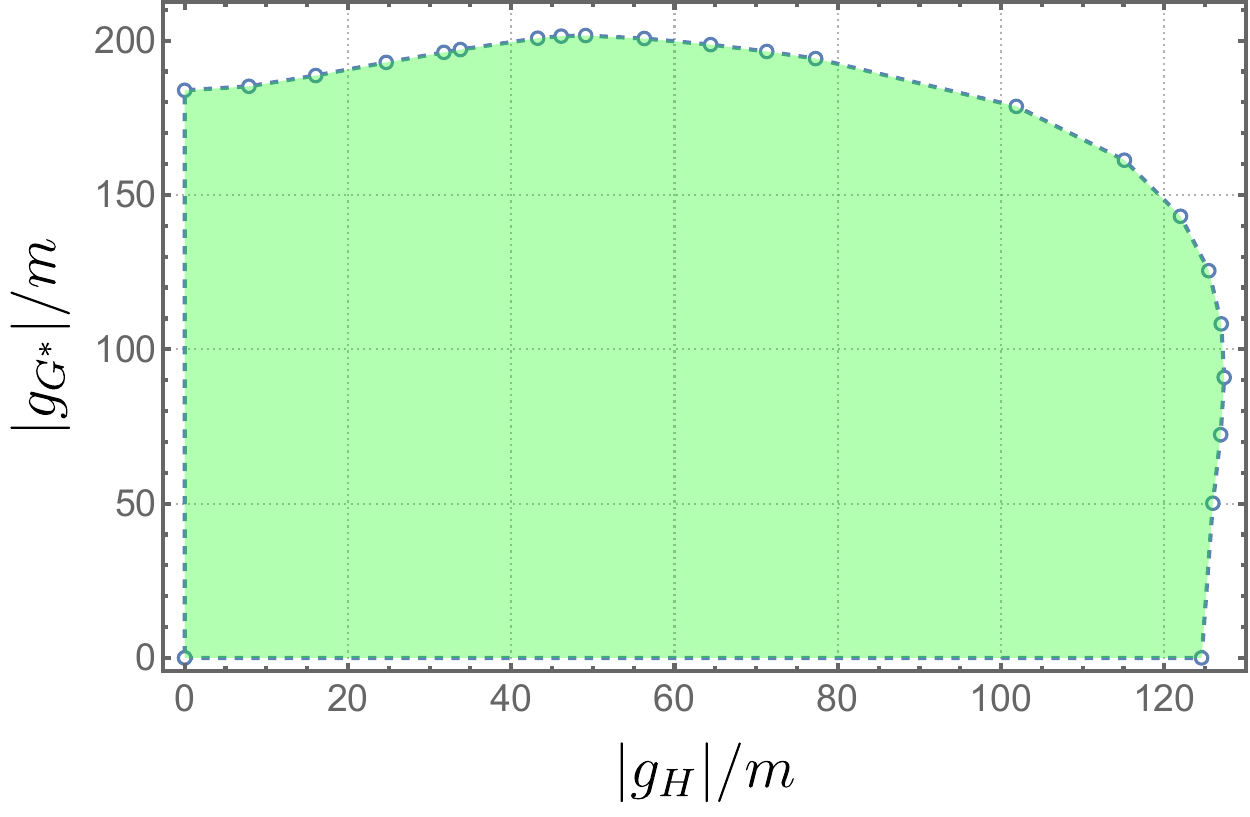}
 	\caption{2D bounds on the coupling constants of the glueballs. The bound in the left plot is agnostic about $g_{G^*}$, while that in the right plot is agnostic about $g_{G}$. The numerical parameters are $\ell_{\rm max} = 24$, $k_{\rm max} = k^{(H)}_{\rm max} = 32$ and $s_{\rm max} =12 m^2$.} 
 	\label{fig:glueball}
\end{figure}

Figure \ref{fig:glueball} shows the 2D bound on $g_H$ and $g_G$ while agnostic about $g_{G^*}$, and the 2D bound on $g_H$ and $g_{G^*}$ while agnostic about $g_{G}$. Since the reported numerical uncertainties of the dual bounds in \cite{Guerrieri:2023qbg} are still sizable, we limit ourselves to a rough visual comparison, which shows reasonably good agreement, hence providing strong confidence in the employed methods.

\section{Fractionally subtracted dispersion relations}
\label{sec:fractional}

As mentioned, our primal bootstrap approach has the advantage of taking into account the Jin-Martin bound via the twice-subtracted dispersion relation. A natural question to ask is what is the consistent theory space for models where the amplitude grows slower than $|s|^2$ at large $|s|$? Do we need the amplitude to saturate the Jin-Martin bound to reach the boundary of the theory space? To address these questions systematically, we shall introduce a dispersion relation with a fractional subtraction order. Fractionally subtracted dispersion relations were previously used in \cite{deRham:2022gfe} to derive continuous moment sum rules, see also \cite{Liu:1967xaf,Olsson:1968fuu}.

\subsection{Dispersion relation with fractional subtraction order}

We shall return to the case of a scalar field theory with mass $m$ and without bound states below the threshold. Let us consider a theory where the amplitude has the following large $|s|$ behavior 
\begin{equation}
    \lim_{|s|\to\infty}|\mc{M}(s,t)|<|s|^{2r}\,,~~~~0\leq r\leq 1 \,,
\end{equation}
for $-4m^2<t<4m^2$ and $r$ being a constant. The derivation of a dispersion relation hinges on the vanishing of the infinite arc contributions in the $s$-plane contour, for which we must ``subtract'' to soften the amplitude’s asymptotic behavior. To that end, we can multiply the amplitude with the following $su$ crossing-symmetric kernel
\begin{equation}
\label{kernalGrs}
    G^{\{r_i\}}_{\{s_i\}}(s,t) = \prod_{i=1}^n (s_i-s)^{-r_i}(s_i-4m^2+s + t)^{-r_i}\,,
\end{equation}
where we choose constants $r_i <1$ to avoid introducing poles
\begin{equation}
    4m^2=s_0<s_1<s_2<\dots<s_n<s_{n+1}=\infty\,,~~~~
 0<r_i<1,~~\sum_{i=1}^n r_i = r\,.
\end{equation}
In this paper, we shall work with the simplest choice of two subtraction points, $s_1$ and $s_2$ (with $n=2$). We will also choose $r_1$ and $r_2$ to be smaller than $1/2$, which improves numerical convergence for the (time-consuming) integrations near the branch points. Note that, for fixed $t$, the analytical structure of the kernel is simple: it is analytic on the complex $s$-plane except for two branch cuts $s\geq s_1$ and $s \leq 4m^2-s_1-t$, which crucially ``hide'' behind the physical branch cuts of the amplitude. Multiplied by this kernel, the modified amplitude now has the following asymptotical behavior
\begin{equation}
    \lim_{|s|\to \infty}\Big|\mc{M}(s,t)G^{\{r_i\}}_{\{s_i\}}(s,t)\Big| < {\rm const}\,, ~~~ -4m^2<{\rm fixed~}t<4m^2 \,.
\end{equation}
Hence, by Cauchy's integral formula and $su$ crossing symmetry,  we immediately get 
\begin{equation}
    \mc{M}(s,t)G^{\{r_i\}}_{\{s_i\}}(s,t) = \int_{4m^2}^{\infty}\frac{\d \mu}{\pi} {\rm Disc}_{\mu}\left(\mc{M}(\mu,t)G^{\{r_i\}}_{\{s_i\}}(\mu,t) \right) \left(\frac{1}{\mu-s}+\frac{1}{\mu-4m^2+s+t}\right)\,,
\end{equation}
where the discontinuity can be written as 
\begin{equation}
\label{discM2r}
\begin{aligned}
    {\rm Disc}_{\mu}\left(\mc{M}(\mu,t)G^{\{r_i\}}_{\{s_i\}}(\mu,t) \right)& = \left(\sin{\psi_j} \, \re \mc{M}(\mu,t) + \cos{\psi_j}\, \im \mc{M}(\mu,t)\right) \Big|G^{\{r_i\}}_{\{s_i\}}(\mu,t) \Big|,\\&\text{ $\psi_j = \sum_{i=1}^{j}r_j\pi$~~ for~~ $s_j< \mu<s_{j+1}$, }
\end{aligned}
\end{equation}
thanks to real analyticity of the amplitude. Similar to the twice-subtracted case, the imaginary part of the above dispersion relation is again trivial, while for $s> 4m^2$ and $-4m^2<t<4m^2$, its real part can be written as,   
\begin{equation}
\begin{aligned}
    \re \left( \mc{M}(s,t)G^{\{r_i\}}_{\{s_i\}}(s,t)\right)  = \mc{P}\int_{4m^2}^{\infty}\frac{\d \mu}{\pi} {\rm Disc}_{\mu}&\left(\mc{M}(\mu,t)G^{\{r_i\}}_{\{s_i\}}(\mu,t) \right) \left(\frac{1}{\mu-s}+\frac{1}{\mu-4m^2+s+t}\right)\,,
\end{aligned}
\end{equation}
where the left hand side of the above equation can be written as
\begin{equation}
    \begin{aligned}
    \re \left( \mc{M}(s,t)G^{\{r_i\}}_{\{s_i\}}(s,t)\right)& = \left(\cos{\psi_j} \, \re \mc{M}(s,t) - \sin{\psi_j}\, \im \mc{M}(s,t)\right) \Big|G^{\{r_i\}}_{\{s_i\}}(s,t) \Big|,\\&\text{ $\psi_j = \sum_{i=1}^{j}r_j\pi$~~ for~~ $s_j< s<s_{j+1}$. }
\end{aligned}
\end{equation}
This $2r$-th subtracted dispersion relation is the one we will use for the numerical bootstrap in this section. Note that the dispersion relation with $2r = 2$ in this setup differs slightly from the standard twice-subtracted form; however, as we will see, both yield the same constraints on theory space. 

Note that the kernel \eqref{kernalGrs} separates the $4m^2 \le \mu <\infty $ region into several sub-regions. An advantage of this separation is that we can parametrize the partial wave amplitudes in these different sub-regions independently, with each sub-region approximated by continuous polynomials, which thus allows for discontinuities at the separation points. This is a welcome feature physically. For example, for a single scalar theory, there are many branch points above the threshold $4m^2$ at $\mu=9m^2, 16m^2,...$, corresponding to various channels of higher multi-particle production at loop levels. To take into account this scenario, we can set the separation points at those values of $\mu$.

\subsection{Bounds on $g_0$ and $g_2$ again}
\label{sec:2rsubg0g2}

Having established the $2r$-th subtracted dispersion relation, we shall now adjust the previous bootstrap approach to compute numerical bounds. For the twice-subtracted dispersion relation, we parametrize the discontinuity within the dispersive integral, up on partial wave expansion, that is, we parameterize the imaginary parts of the partial waves. For the $2r$-th subtracted dispersion relation, the discontinuity now contains both the real and imaginary parts of the partial waves. In the current case, it is convenient to parametrize the partial waves of the (whole) discontinuity in the dispersive integral, that is, we parameterize the following expressions, 
\begin{equation}
   \sin{\psi_j}\, \re\, a_{\ell}(\mu) + \cos{\psi_j}\, \im\, a_{\ell}(\mu)\text{, $~~~\psi_j = \sum_{i=1}^{j}r_j\pi$~~ for~~ $s_j< \mu<s_{j+1}$, }
\end{equation}
for $\mu\geq 4m^2$. With these expressions parameterized, we then use the same method outlined around \eref{Re_a_ellSol} to compute $\cos{\psi_j}  \re\, a_{\ell}(s) - \sin{\psi_j} \im\, a_\ell(s)$, which then fixes the real and imaginary parts of the partial wave amplitudes.

Note that our discussions on the extra factors in the partial wave ansatz are still valid for this linear combination of the real and imaginary part of the partial wave\,\footnote{Near the threshold $4m^2$, the discontinuity is still $\im\mc{M}(\mu,t)$, so the previous discussion on the threshold behavior still applies.}. So the extra factors that go into the ansatz can still be chosen as 
\begin{equation}
    \sqrt{\frac{\mu}{\mu-4m^2}}\left(\frac{\mu-4m^2}{\mu+4m^2+4m\sqrt{\mu}}\right)^{\ell}\,.
\end{equation}
The difference is that, now, we need to choose a separate functional basis for each segment $s_i<\mu<s_{i+1}$. For $s>s_n$, we will use the functional basis $P_k(2s_n/\mu-1)$, and for other ranges, we can choose $P_k(2(\mu-s_i)/(s_{i+1}-s_{i})-1)$. The maximal order of these functional bases will be denoted by $k^{(0)}_{\rm max},k^{(1)}_{\rm max},\dots, k^{(n)}_{{\rm max}}$. Note that increasing the order of one of the $k^{(m)}_{{\rm max}}$'s often necessitates adding more discrete points of $\mu$ in all regions, as $\cos{\psi_j}  \re\, a_{\ell}(s) - \sin{\psi_j} \im\, a_\ell(s)$ on the left hand side of the dispersion relation are computed from the expressions with the dispersive integrations carried out. Another minor difference is that, for the discrete points above $s_{\rm max}$ where we impose linear unitarity, we now use the same sample points as those used in the numerical integration. The reason for this is that the integration for each partial wave converges more slowly when using the $2r$-th subtracted dispersion relations with small $r$. Therefore, it is preferable to impose the linear unitarity condition at larger values of $s$.

Aside from these subtleties, our primal bootstrap setup remains largely unchanged for the fractionally subtracted dispersion relation. The $2r$-th subtracted dispersion relation provides us a tool to probe how the bounds depend on the Regge behavior of the amplitudes.

\begin{figure}
    \centering
    \includegraphics[width=0.6\linewidth]{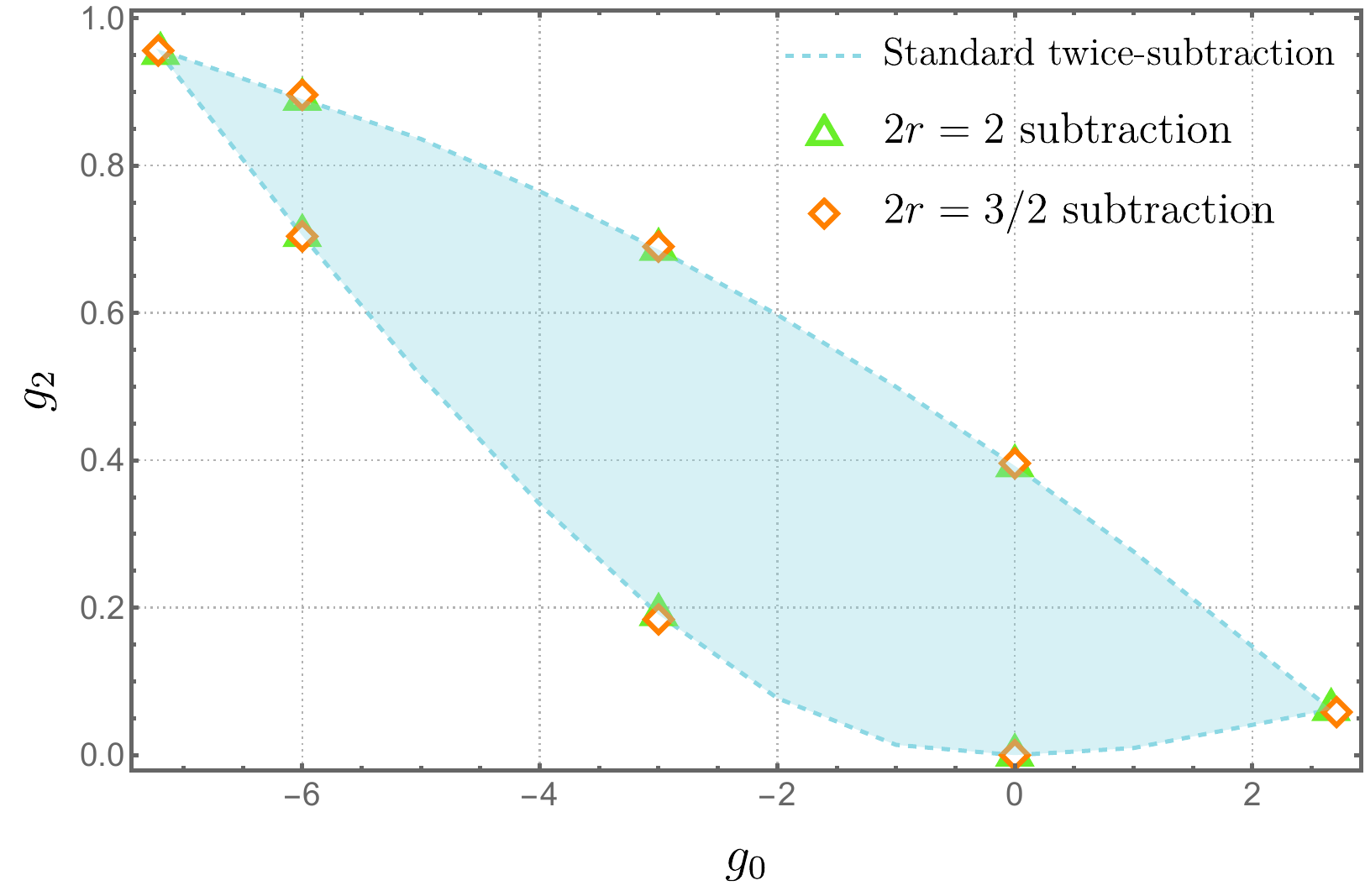}
    \caption{Bounds on $g_0$ and $g_2$ from a 2nd subtracted and $3/2$-th subtracted dispersion relation, as compared to those from the standard twice-subtracted one ({\it i.e.,} from Figure \ref{fig:2d_g0g2}). The $r$ parameters are $r_1=r_2=1/2$ for the $2$nd subtracted dispersion relation, and $r_1=r_2=3/8$ for the $3/2$-th subtracted dispersion relation. For the upper-bound kink, we use $\ell_{\rm max}=24$, while for other marked points we use $\ell_{\rm max}=42$. The rest parameters are chosen as $s_1=9m^2$, $s_2=16m^2$, $s_{\rm max}=72m^2$ and $k^{(0)}_{{\rm max}}=26,~ k^{(1)}_{{\rm max}}=26,~ k^{(2)}_{{\rm max}}=46$.}
    \label{fig:newdisp_1}
\end{figure}

\begin{figure}
    \centering
    \includegraphics[width=0.57\linewidth]{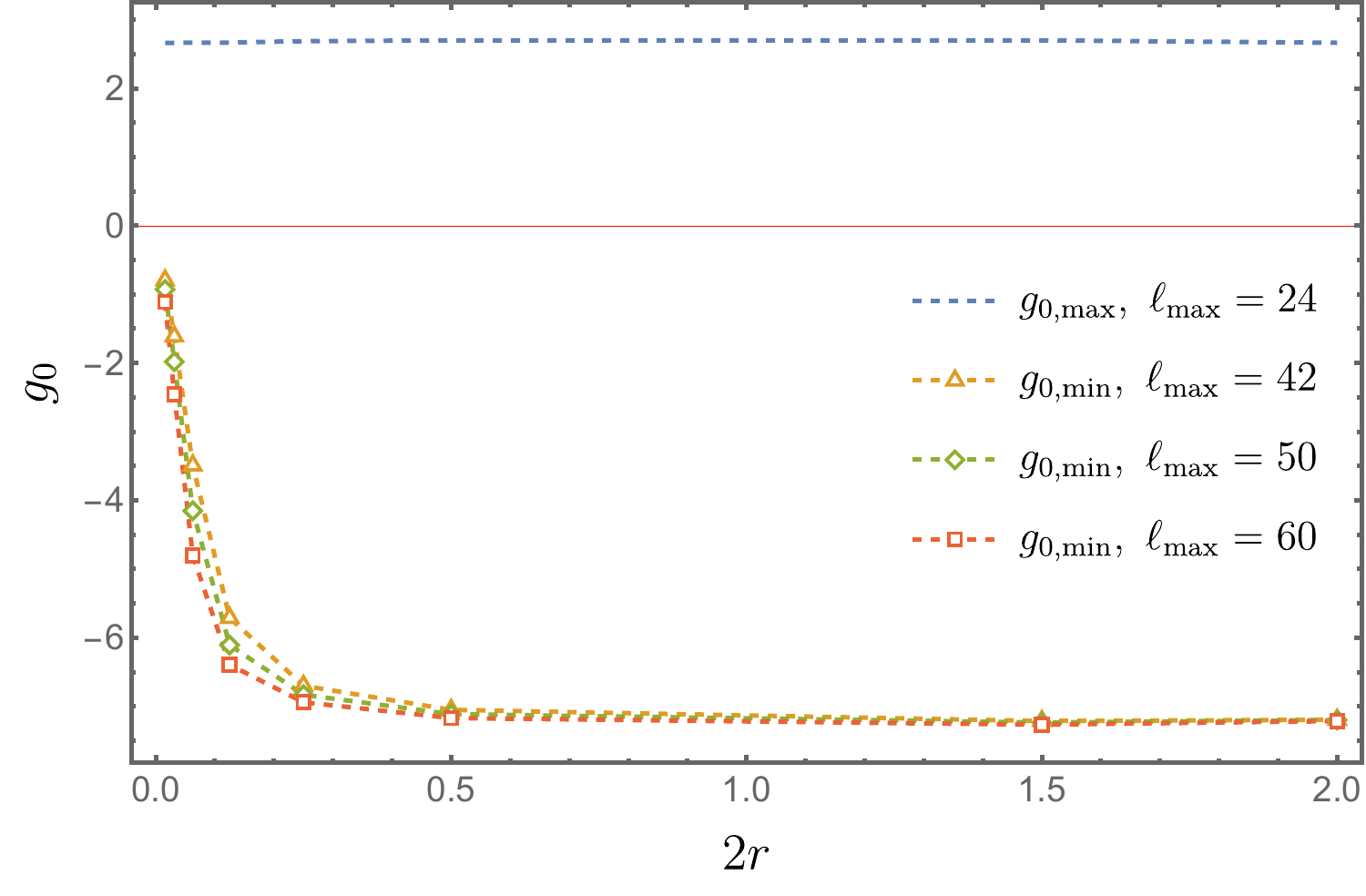}
    \caption{Dependence of the two-sided $g_0$ bounds on  the subtraction order $2r$. For the $2$nd subtracted and $3/2$-th subtracted dispersion relations, we use $s_1=9m^2,~s_2=16m^2$ while for other dispersion relations we use $s_1=10m^2,~s_2=16m^2$. The rest parameters are chosen as $r_1=r_2$, $s_{\rm max}=72m^2$ and $k^{(0)}_{{\rm max}}=26,~ k^{(1)}_{{\rm max}}=26,~ k^{(2)}_{{\rm max}}=46$.}
    \label{fig:newdisp_2}
\end{figure}

In Figure \ref{fig:newdisp_1}, we see that the bound on $g_0$ and $g_2$ is almost the same if we lower the subtraction order from $2$nd to $3/2$-th. This is more or less consistent with the results in Section \ref{sec:regge}, where we saw that the fastest asymptotical growth rate of the amplitude is around $s^{1.5}$ or slightly faster near $t=4m^2$. (The impact of varying the integer subtraction order on the positivity bounds has previously been investigated by imposing the positivity of partial wave unitarity \cite{McPeak:2023wmq}.) We can also see from this figure that at least the bounds themselves are insensitive to the locations of the partial wave discontinuities $s_i$.

Furthermore, Figure \ref{fig:newdisp_2} shows that the upper bound on $g_0$ is unchanged as the subtraction order $2r$ goes to zero. This is not surprising as we have already seen in Section \ref{sec:bound_expansion_coeff}, the amplitude at the upper bound of $g_0$ asymptotes to a constant (or slowly decays) at large $|s|$, hence the insensitivity to the subtraction order. For the lower bound on $g_0$, however, the bound starts to shrink when the subtraction order $2r$ goes below around $0.5$. Also, we see that as $2r$ goes to zero, the lower bound on $g_0$ approaches zero. This is what one would expect:  For a non-subtracted dispersion relation, the coefficient $g_0$ becomes dispersive, that is, we have the following positive sum rule
\begin{equation}
    g_0 = \frac{1}{16\pi^2}\int_{4m^2}^{\infty} \d\mu \frac{\im \mc{M}(\mu,4m^2/3)}{\mu-4m^2/3} \geq 0\,,
\end{equation}
where the last inequality comes from the unitary condition $\im \mc{M}(\mu,4m^2/3)\geq 0$ for $\mu\geq 4m^2$. An interesting observation that can be inferred from Figure \ref{fig:newdisp_2} is that most of the parameter space in the negative $g_0$ region is consistent with very slow Regge behavior $\mc{M}\lesssim s^{0.1}$, which in itself may provide valuable insight on connections with the high energy behavior.

\section{Conclusion}
\label{sec:conclu}

In this work, we have unveiled and demonstrated the benefits of
a new framework for systematically constructing the consistent, non-perturbative amplitude space of quantum field theories. 
Using a scalar theory as an illustrative example, we have demonstrated the effectiveness of our bootstrap method. Our results are consistent with those obtained from previous methods where applicable, while also showing noticeable improvements in both physical and numerical aspects. Owing to the integral use of the dispersion relation, our approach allows for the straightforward inclusion of bound states with spin below the threshold, achieving excellent numerical convergence. Leveraging this feature, we have applied our formalism to constrain realistic glueball coupling constants, using lattice QCD spectral data as input.

Our approach relies solely on the rigorously established analyticity domain of Martin and parameterizes the partial wave amplitudes entirely within the physical region, thereby remaining closer to directly observable quantities. This contrasts with the earlier primal approach, which requires maximal analyticity and constructs amplitudes in analytically continued regions using more abstract variables. A key advantage of our method is that it incorporates the asymptotic high-energy behavior of the amplitude by construction, enabling us to build the theory space from a complementary perspective. Note that the maximal growth rate that can be constructed is tantalizingly close to the proven Jin-Martin upper bound, though a small gap to $s^2$ remains. It would be interesting to determine whether this gap reflects a genuine theoretical obstruction or merely stems from numerical inefficiency.

The use of dispersion relations provides direct control over the asymptotic growth, allowing us to probe the sensitivity of the theory space to different high-energy behaviors. This is accomplished through the introduction of a new class of dispersion relations with fractional subtraction orders. These fractionally subtracted dispersion relations naturally partition the unitarity cut above the threshold into multiple regions, facilitating the incorporation of multi-particle thresholds in the numerical bootstrap and the probe of the effects of them.

For future extensions, it is worth noting that a stronger set of constraints yet to be imposed is elastic partial wave unitarity below the second multi-particle threshold, where the unitarity bounds are saturated as equalities. However, these constraints are non-convex and therefore more challenging
to implement in the current numerical setup. Also, we only use the Martin's analyticity range for $t$ for a generic massive scalar, which is $-4m^2 < t< 4m^2$ in the absence of bound states below the threshold. This analyticity range can be further extended in specific theories. For instance, in the case of pion-pion scattering, it extends to $-28m^2 < t < 4m^2$ \cite{Martin:1965jj}. However, a preliminary study using this extended range does not appear to enlarge the space of consistent theories, at least with respect to the leading amplitude coefficients. Further investigation is, of course, warranted. Another extension of this primal method is its application to the EFT framework. For a weakly coupled EFT, the absence of threshold singularities significantly simplifies the numerical implementation. As a result, we observe notable improvements in the EFT bounds compared to those obtained using the previous primal approach \cite{EFTpaper}. Finally, by construction, our bootstrap method is readily generalizable to more complex theories. We find it compelling to explore applications of our method to more complex phenomenological models, including the Standard Model EFT.

\section*{Acknowledgments}

We would like to thank Andrea Guerrieri, Dong-Yu Hong, Martin Kruczenski, Yue-Zhou Li, Balt C. van Rees, Shi-Lin Wan and Alexander V. Zhiboedov for helpful discussions. SYZ acknowledges support from the National Natural Science Foundation of China under grant No.~12475074, No.~12075233 and No.~12247103. The work of CdR and AJT is supported by STFC Consolidated Grant ST/X000575/1. CdR is also supported by a Simons Investigator award 690508. This research is also supported by the advanced computing resources provided by the Supercomputing Center of the USTC.\\

\appendix

\section{Threshold behavior of $\im\, a_{\ell}(s)$}
\label{sec:thresholdOfa_ell}

In Section \ref{sec:paramPartWaves}, we used the fact that $\im\, a_{\ell}(s)$ must decrease faster than $(s-4m^2)^{\ell-1}$ near the threshold. In reaching this statement, we implicitly used the assumption that we can exchange the order of the integration over $s$ and the summation over $\ell$. 
Naively, there is a possibility that some partial waves might not fall off faster than $(s-4m^2)^{\ell-1}$ near the threshold, but still the summation over $\ell$ remains convergent due to possible cancellations, bearing in mind that the $\ell$ summation should, in principle, be performed first. In this appendix, we see that this possibility is excluded by unitarity. 

Note that we are interested in the threshold behavior of $\im\, a_{\ell}(s)$, which is independent of $t$ and $(s_0,t_0)$ in the dispersion relation, so we will choose {\it ad hoc} values of them for our purpose here: $(s_0,t_0)=(2m^2,0)$ and $0<t<4m^2$. Let us focus on the first term of the integral on the right hand side of the main dispersion relation \eqref{imtoreal}, specifically looking at the part near the threshold, 
\begin{equation}
\label{AppIntMsp}
    \int_{4m^2}^{s_1} \f{\d \mu}{\pi} ~ \im\mc{M}(\mu,t)K^{\mu,t_0}_{s,t}\,,
\end{equation}
where we choose $s>s_1>4m^2$ and the principle value evaluation $\mc{P}$ can be neglected for these choices of parameters. The partial wave expansion gives
\begin{equation}
    \im \mc{M}(\mu,t) = 16\pi\sum_{\ell,{\rm even}}(2\ell+1)\im\, a_{\ell}(\mu)P_{\ell} \left(1+\frac{2t}{\mu-4m^2}\right)\,.
\end{equation}
The integral \eqref{AppIntMsp} is convergent because the dispersion relation \eqref{imtoreal} is convergent and the second term on the right hand side of \eqref{imtoreal} is convergent due to the fact that $P_{\ell}(1 + 2t_0/(\mu-4m^2))|_{t_0 = 0}$ does not diverge at $\mu = 4m^2$. The key point is that unitarity tells us $\im\, a_{\ell}(\mu) \geq 0$ for $\mu\geq 4m^2$, and $P_{\ell}(1+2t/(\mu-4m^2))$ is positive since we have chosen a positive $t$. Therefore, we have 
\begin{equation}
    \im \mc{M}(\mu,t)\geq 16\pi (2\ell_0+1)\im  a_{\ell_0}(\mu) P_{\ell_0}\left( 1 + \frac{2t}{\mu -4m^2}\right)\,,
\end{equation}
for any fixed even $\ell_0$. Also, thanks to the choice $t_0=0$, we have $K^{\mu,t_0}_{s,t}<0$ for $4m^2 <\mu<s_1$. Therefore we have 
\begin{equation}
    \bigg|\int_{4m^2}^{s_1} \f{\d \mu}{\pi}~\im\mc{M}(\mu,t)K^{\mu,t_0}_{s,t}\bigg|\geq 16\pi (2\ell_0 +1)\bigg| \int_{4m^2}^{s_1} \f{\d \mu}{\pi}~ \im\, a_{\ell}(\mu) P_{\ell_0}\left( 1 + \frac{2t}{\mu -4m^2}\right)K^{\mu,t_0}_{s,t}\bigg|\,,
\end{equation}
which means that if one partial wave leads to divergence, the dispersion relation will diverge. Since the dispersion relation is convergent, we must have
\begin{equation}
    \lim_{s\to 4m^2}\frac{\im\, a_{\ell}(s)}{(s-4m^2)^{\ell-1}} = 0 \,.
\end{equation}

\section{Large $\ell$ behavior of partial waves} 
\label{sec:largel}

In this appendix, we review the large $\ell$ behavior of the partial wave amplitudes. We shall start with the Froissart-Gribov formula in $d$-dimensional spacetime:
\begin{equation}
\label{FGformula}
	a_{\ell}(s)=2\mc{N}_d \int_{z_1(s)}^{\infty} \frac{\d z}{\pi} (z^2-1)^{\frac{d-4}{2}} Q_\ell^{(d)}(z){\rm Disc}_z \mc{M}(s,t(z))\,,
\end{equation}
where $t=(s-4m^2)(z-1)/2$, $z_1= 1+8m^2/(s-4m^2)$, $\mc{N}_d=(16\pi)^{(2-d)/2}/\Gamma((d-2)/2)$ and $Q^{(d)}_\ell (z)$ is the Gegenbauer $Q$ function
\begin{equation}
	Q^{(d)}_\ell (z)=\frac{C_{\ell}^{(d)}}{z^{\ell+d-3}} {}_{2}F_{1} \Big(\frac{\ell+d-3}{2},\frac{\ell+d-2}{2},\ell+\frac{d-1}{2},\frac{1}{z^2}\Big)\,,
\end{equation}
with $C_{\ell}^{(d)}={\sqrt{\pi}\Gamma(\ell+1)\Gamma(\frac{d-2}{2})}/({2^{\ell+1}\Gamma(\ell+\frac{d-1}{2})})$. This formula can be derived by combining the partial wave expansion with the fixed-$s$ dispersion relation arising from maximal analyticity, and is valid for a sufficient large $\ell$ that depends on the polynomial boundedness of the amplitude.

Note that, for large $\ell$, $Q_{\ell}^{(d)}$ has the following asymptotic behavior
\begin{equation}\label{largel_Q}
	Q_\ell^{(d)}(z)= 2^{d-4}\sqrt{\pi}\frac{\Gamma (\frac{d-2}{2})}{\ell^{\frac{d-3}{2}}}\frac{(\lambda(z))^{-\ell}}{(\lambda(z)^2-1)^{\frac{d-3}{2}}}(1+\mc{O}(1/\ell))\,,
\end{equation}
where $\lambda(z)=z+\sqrt{z^2-1}$. Thus, when $\ell$ is large, $Q^{(d)}_{\ell}(z)$ falls off very fast with $\ell$ for $z>1$, and the integration in the Froissart-Gribov formula is dominated by the region near the lower limit $z\sim z_1(s)$. We can then expand \eref{largel_Q} around $z=z_1(s)$:
\begin{equation}
	Q_\ell^{(d)}(z_1+\delta z)=2^{d-4}\sqrt{\pi}\frac{\Gamma (\frac{d-2}{2})}{\ell^{\frac{d-3}{2}}}\frac{(\lambda(z_1))^{-\ell}}{(\lambda(z_1)^2-1)^{\frac{d-3}{2}}} \exp\bigg(-\frac{\ell \delta z}{\sqrt{z_1^2-1}}\bigg)\,.
\end{equation}
On the other hand, for a fixed $\ell_{0}$, we have the following inequality for $\ell\geq \ell_0$
\begin{equation}
	\begin{aligned}
		&~~~~\bigg|\int_{0}^{\infty} \d\delta z {\rm Disc}_t \mc{M}(s,t=(s-4m^2)(z_1+\delta z-1)/2)\exp\bigg(-\frac{\ell \delta z}{\sqrt{z_1^2-1}}\bigg)\bigg|
		\\&\leq \int_{0}^{\infty} \d\delta z \bigg|{\rm Disc}_t \mc{M}(s,t=(s-4m^2)(z_1+\delta z-1)/2)\bigg|\exp\bigg(-\frac{\ell_0 \delta z}{\sqrt{z_1^2-1}}\bigg)\,,
	\end{aligned}
\end{equation}
where the right hand side converges thanks to the polynomial boundedness of the amplitude. Substituting these back into the Froissart-Gribov formula, as $\ell$ increases, we see that at fixed $s$, the large-spin partial wave amplitude decays exponentially with $\ell$
\begin{equation}
	|a_{\ell}(s)|\leq g^{(d)}(s) \frac{1}{\ell^{\frac{d-3}{2}}} \bigg(\frac{s-4m^2}{s+4m^2+4m \sqrt{s}}\bigg)^\ell\text{,~~~ for sufficiently large $\ell$}, \label{large_ell_partialwave}
\end{equation} 
where $g^{(d)}(s)$ is a positive function of $s$. 

Generally, the high-spin partial waves become more important when $s$ increases, as can be seen from the expressions above. For example, as $s$ increases, the fraction in the bracket of \eref{large_ell_partialwave} approaches 1. Thus, more partial waves are needed to approximate the amplitude for a larger $s$.

\section{Numerical scheme for dispersive integrals}\label{sec:cal_int}

As mentioned, efficient numerical evaluation of the dispersive integral to a high precision (up to $\mc{O}(1000)$ digits) is crucial for achieving both high accuracy and fast speed in our method. In this appendix, we specify the integration scheme we use.

Recall that in our method we need to calculate the principal value of the following integral
\begin{equation}
    I(s,t):=\mc{P}\left(\int_{4m^2}^{\infty} \d \mu ~ \im\mc{M}(\mu,t) K^{\mu,t_0}_{s,t} +  \int_{4m^2}^{\infty} \d \mu ~ \im\mc{M}(\mu,t_0) K^{\mu,s_0}_{t,t_0}\right)\,.
\end{equation}
This integral has improper points at $\mu = 4m^2$, $\mu = s$ and potentially $\mu = 4m^2-t-t_0$ if $t< - t_0$. Our strategy is to approximate the regions near the improper points $\mu = s$ and $\mu = 4m^2-t-t_0$ with simple integrals that can be integrated analytically, while the remaining improper point $\mu = 4m^2$ is handled through a suitable change of the integration variable.

The part of $I(s,t)$ near an improper point $\mu = x$ can be approximated by the corresponding part of
\be
I_0(y,t)=\theta(y-4m^2)\mc{P}\int_{4m^2}^{\infty} \d \mu ~\im \mc{M}(y,t) \frac{y}{\mu}\frac{1}{\mu-y},
\ee
where the $y/\mu$ factor is introduced so that the integral is convergent. $I_0(x,t)$ can be evaluated analytically once the ansatz for $\im \mc{M}(\mu,t)$ is given. Canceling the improper points with such integrals, we can define a subtracted integral 
\begin{equation}
   \tilde{I}(s,t)  =  I(s,t) - I_0(s,t) + I_0(4m^2-t-t_0,t)-I_0(4m^2-t-t_0,t_0)  \,, 
\end{equation}
which only has an improper point at $\mu=4m^2$. Once $\tilde{I}(s,t)$ is evaluated, we can obtain the target integal $I(s,t)$ by adding the $I_0$ terms.

To evaluate $\tilde{I}(s,t)$, we first make a change of variable $\mu\to 8m^2/(x+1)$, which changes the integration region to $(-1,1)$. Denoting the integrand of $\tilde{I}(s,t)$ as $\tilde{\mc{I}}(\mu;s,t)$, we then have
\begin{equation}
    \tilde{I}(s,t) = \int_{-1}^{1} \d x ~ \frac{8m^2}{(x+1)^2} \tilde{\mc{I}}\(\mu = \frac{8m^2}{x+1}; s,t\)
\end{equation}
A widely used numerical technique for high-precision integration over the interval $(-1,1)$ is the tanh-sinh quadrature method. To apply this method, we change the variable via $x = \tanh\left(\frac{\pi}{2} \sinh w\right)$, under which the integral becomes
\begin{equation}
	\begin{aligned}
		 \tilde{I}(s,t) &=\int_{-\infty}^{\infty}\d w   \cosh w \,\bigg[4\pi m^2 \frac{1-x}{1+x} \,  \tilde{\mc{I}}\left(\mu=\frac{8m^2}{x+1};s,t\right)\bigg]\bigg|_{x=\tanh(\pi/2 \sinh w)}
		\\&:=\int_{-\infty}^{\infty}\d w \mc{J}(w;s,t) \,.
	\end{aligned}
\end{equation}

Note that after these transformations, the improper point $\mu=4m^2$ is mapped to $w=\infty$, which is a regular point in the $w$ integration. At this point, we can safely replace the integration as a discrete sum
\begin{equation}
\label{ItildeDis}
	\tilde{I}(s,t)\approx \frac{1}{2^N}\sum_{h=-h_{\rm max}}^{h_{\rm max}} \mc{J}(h/2^N;s,t) \,.
\end{equation}
In practice, we find that a choice of $N=8$ and $h_{\max}\sim \mc{O}(2\times 10^3)$ is sufficient for the precision we need---the precision of the discretized tanh-sinh method scales as $e^{-1/\d w}$. The Legendre polynomials in the integrals are evaluated using the recurrence relation.

In general, for each numerical bootstrap run, we must perform a numerical integration for every decision variable $c_{\ell,k}$ at each discrete point of $s$ and $t$. As a result, the total number of $\tilde{I}$ integrations is $(\frac{\ell_{\rm max}}{2} + 1)^2 \times (k_{\rm max} + 1) \times N_s$, which is computationally expensive. To accelerate the integrations, note that the dependence on $s$ in \eref{ItildeDis} appears only in the kernel $K^{\mu,t_0}_{s,t}$ and the subtraction point $\mu = s$. Therefore, many terms in \eref{ItildeDis} can be computed without sampling $s$. However, these $s$-independent terms generally require higher precision: We use 21,000 bits for their computation, compared to 7,500 bits for the terms involving $K^{\mu,t_0}_{s,t}$. (For the $2r$-th subtracted dispersion relations, higher numerical precision is required, as the integrals of the partial waves converge more slowly for small $r$. In general, this also necessitates using larger values of $s$ as sample points. For example, in the case of the $1/64$-th subtracted dispersion relation, we use 210,000-digit precision to achieve the desired accuracy in the integrations.) Overall, this approach yields a significant numerical speed-up.

The entire procedure is implemented in MPI parallelized C++ code, using the GNU MP and MPFR libraries to enable arbitrary-precision computations. For example, performing the full integration with $\ell_{\rm max} = 82$, $k_{\rm max} = 54$ and $N_s = 1536$ takes approximately 40 hours on a node equipped  with 2 Xeon Platinum 9242 CPUs (96 cores). It is worth emphasizing that this numerical integration only needs to be performed once to get the final ``dispersion relation'' that can be used for various desired optimizations, provided the numerical setup remains unchanged. For smaller values of $\ell_{\rm max}$, $k_{\rm max}$ and $N_s$, it is often sufficient to choose a subset of the data obtained from a higher-precision setup.

\section{Convergence tests} \label{sec:cov_test}

In this appendix, we present some convergence studies supporting the numerical results discussed in the main text.

\begin{figure}
    \centering
    \includegraphics[width=0.57\linewidth]{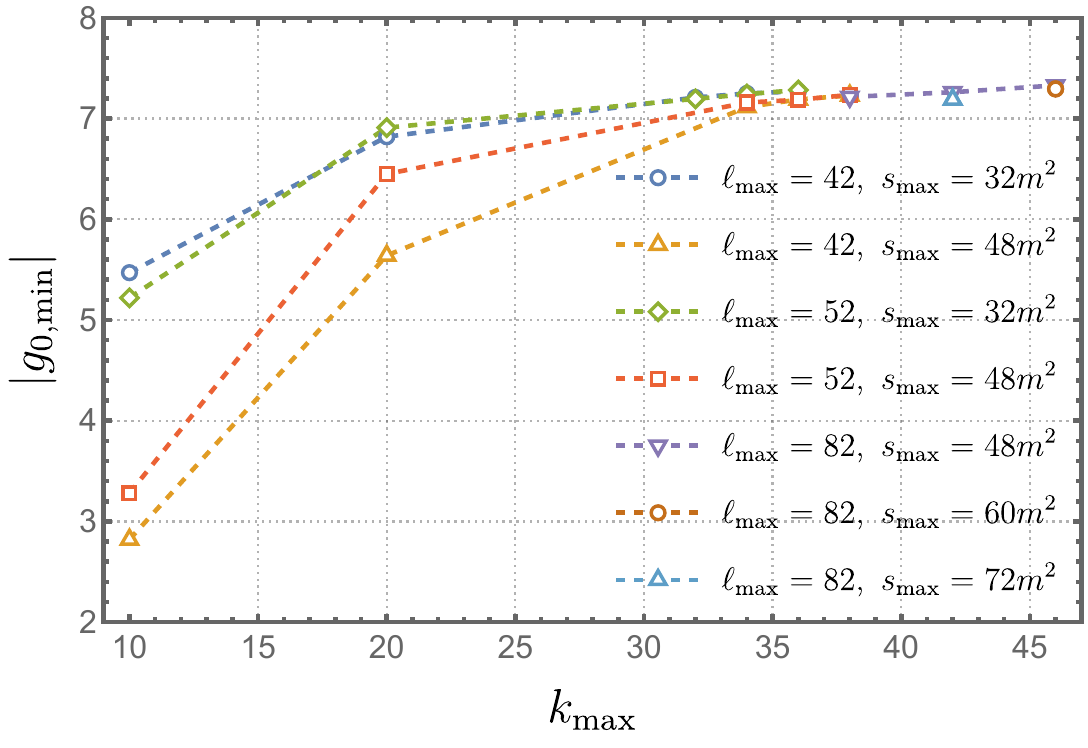}
    \caption{Lower bounds on $g_0$ with different numerical parameters $\ell_{\rm max}, ~s_{\rm max}$ and $k_{\rm max}$. $s$ is discretized according to \eref{eq:disc} with $p=3$. $N_s$ is typically set to be approximately $500$ and $N_s'$ around $1500$, but both should be increased slightly as $k_{\rm max}$ increases. Each chosen pair of $\ell_{\rm max}$ and $s_{\rm max}$ is connected by a dashed line to guide the eye.}
    \label{fig:cov_g0min}
\end{figure}

As mentioned in Section \ref{sec:bound_g0g2}, convergence is generally slower near the minimum of $g_0$. To support the results shown in Figure~\ref{fig:2d_g0g2}, we present in Figure~\ref{fig:cov_g0min} the lower bound on $g_0$ obtained using various numerical parameters: $k_{\rm max}$, $\ell_{\rm max}$, and $s_{\rm max}$. (See Table \ref{tab:numerical_params} for a quick reference of the numerical parameters.). We find that, in general, larger values of $s_{\rm max}$ require correspondingly higher $\ell_{\rm max}$ and $k_{\rm max}$ to ensure good convergence, which is an important consideration to keep in mind when implementing the method. This figure suggests that the numerical parameters used in Figure \ref{fig:2d_g0g2} already lead to good convergence for the lower bound on $g_0$. 

\begin{figure}
    \includegraphics[width=0.5\linewidth]{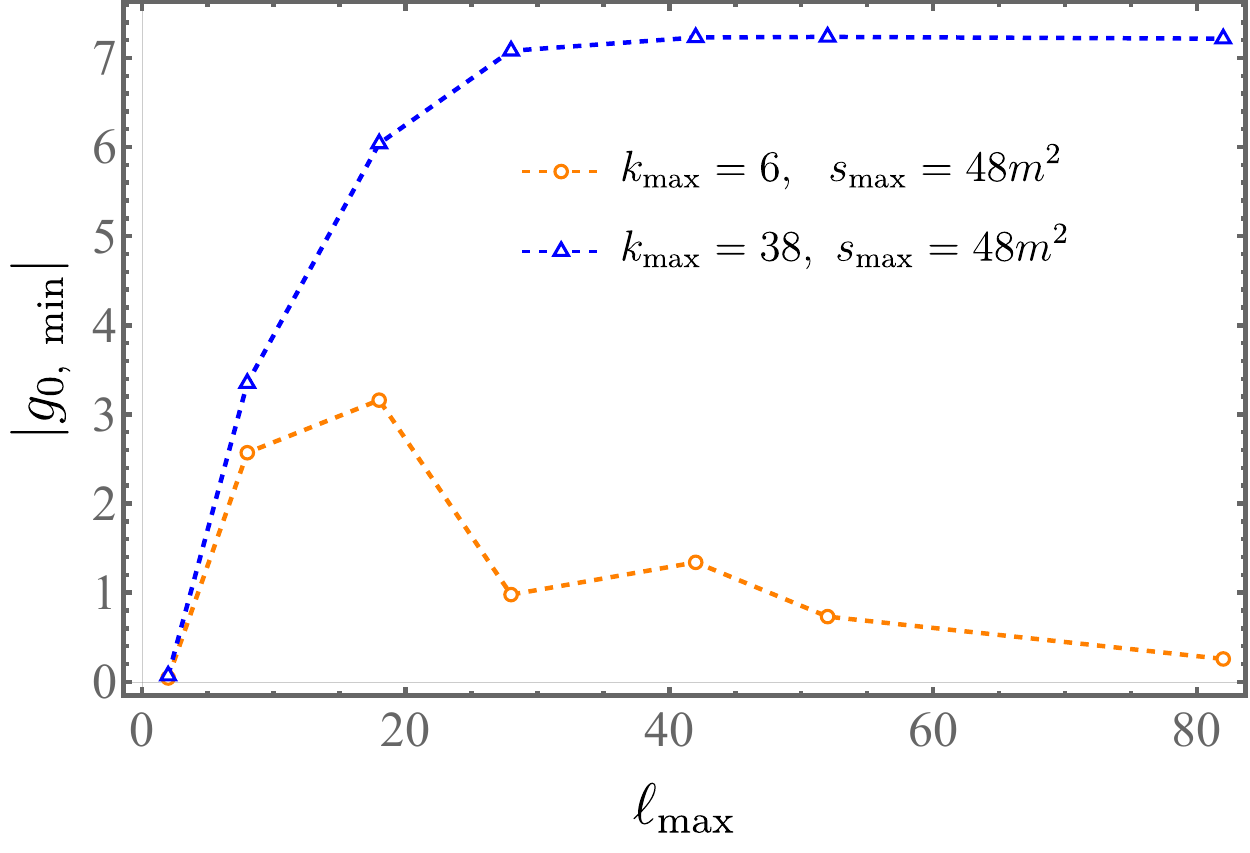}
    \includegraphics[width=0.5\linewidth]{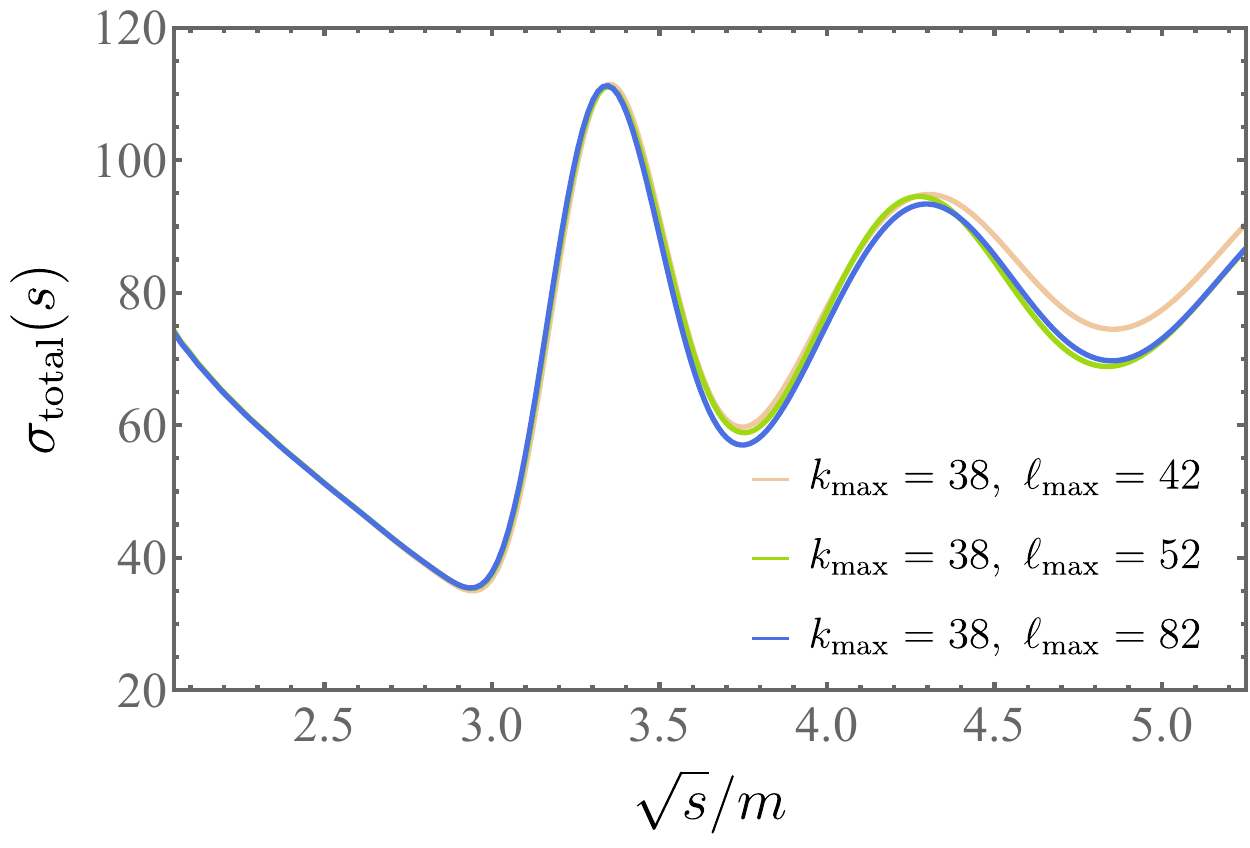}
    \caption{(\textit{Left}) Convergence of the lower bound on $g_0$ with $\ell_{\rm max}$ for different $k_{\rm max}$ values: well-behaved when $k_{\rm max}$ is sufficiently large, and erratic when $k_{\rm max}$ is too small. (\textit{Right}) Total cross section $\sigma_{\rm total}(s)$ at the $g_0$ minimum for different $\ell_{\rm max}$, with $s_{\rm max} = 48m^2$ and $k_{\rm max}=38$, showing good convergence as $\ell_{\rm max}$ increases.}
    \label{fig:cov_sigma}
\end{figure}

Let us also comment on the interplay between the spin cutoff $\ell_{\rm max}$ and the partial-wave basis size $k_{\rm max}$. It is generally observed numerically that once $k_{\rm max}$ is sufficiently large, the convergence with respect to $\ell_{\rm max}$ improves significantly. This is illustrated in Figure~\ref{fig:cov_g0min} for $k_{\rm max}=38$ and $s_{\rm max}=48m^2$, where the three data points corresponding to different $\ell_{\rm max}$ nearly coincide. In cases where $k_{\rm max}$ is not sufficiently large, increasing $\ell_{\rm max}$ may sometimes lead to erratic convergence behavior; such situations can be resolved by increasing $k_{\rm max}$ (see the left plot of Figure~\ref{fig:cov_sigma}). In contract, for a fixed $\ell_{\rm max}$, increasing $k_{\rm max}$ generally improves convergence. To further verify the convergence in partial waves, we plot the total cross section $\sigma_{\rm total}(s) = (s(s-4m^2))^{-1/2}\sum_{\ell} 16\pi (2\ell + 1) \im a_{\ell}(s)$ of the amplitude for these parameter choices in the right plot of Figure~\ref{fig:cov_sigma}.

\begin{figure}
    \centering
    \includegraphics[width=.95\linewidth]{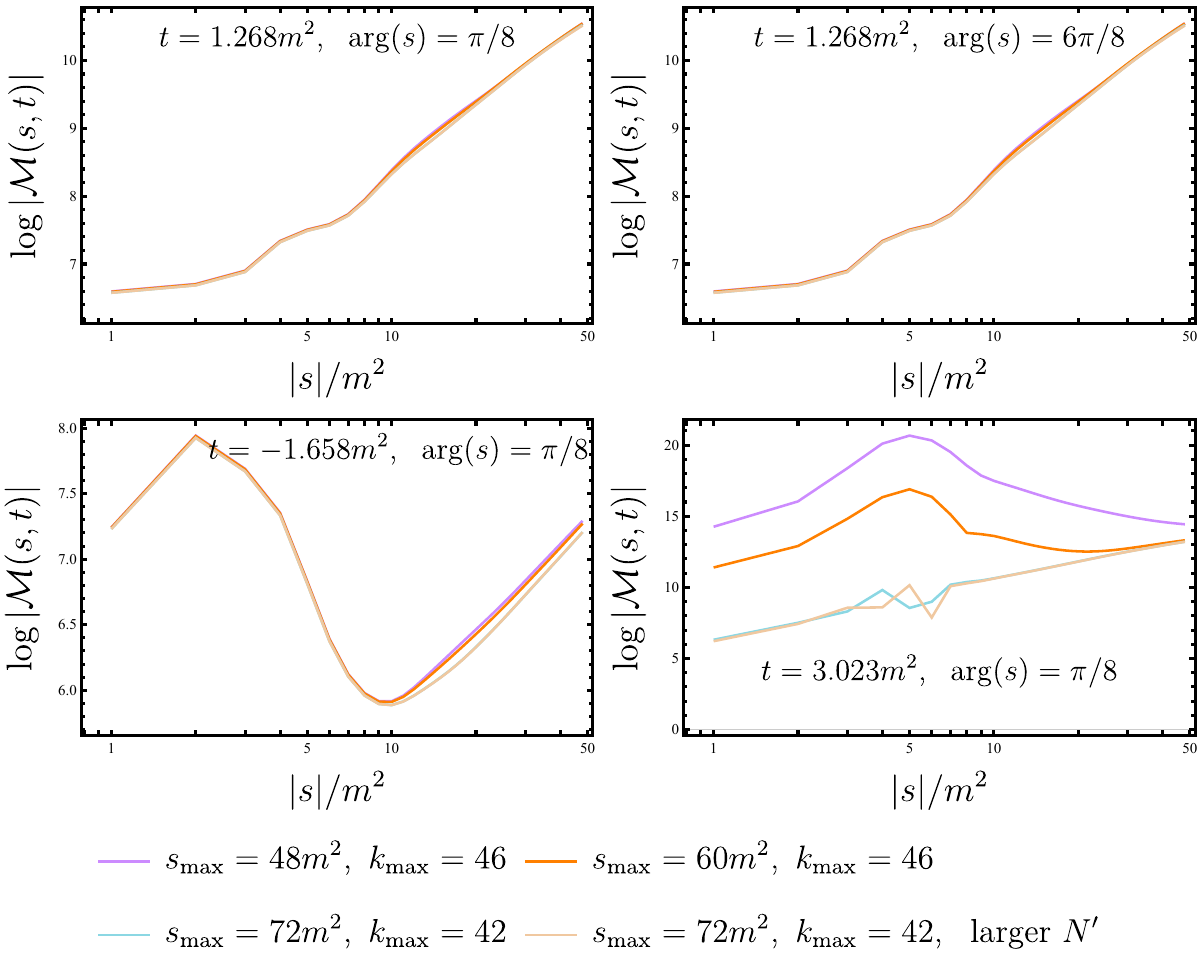}
    \caption{Fixed-$t$ amplitudes at the minimal value of $g_0$ for various $t$ and ${\rm arg}(s)$, with different numerical parameters $s_{\rm max}$ and $k_{\rm max}$. Here $\ell_{\rm max}=82$, $N_s$ is the same as in Figure \ref{fig:cov_g0min}, and we choose $N_s'=2000$ except for the ``large $N'$'' case, where $N_s' = 4000$.}
    \label{fig:cov_amp}
\end{figure}

\begin{figure}
    \centering
    \includegraphics[width=0.55\linewidth]{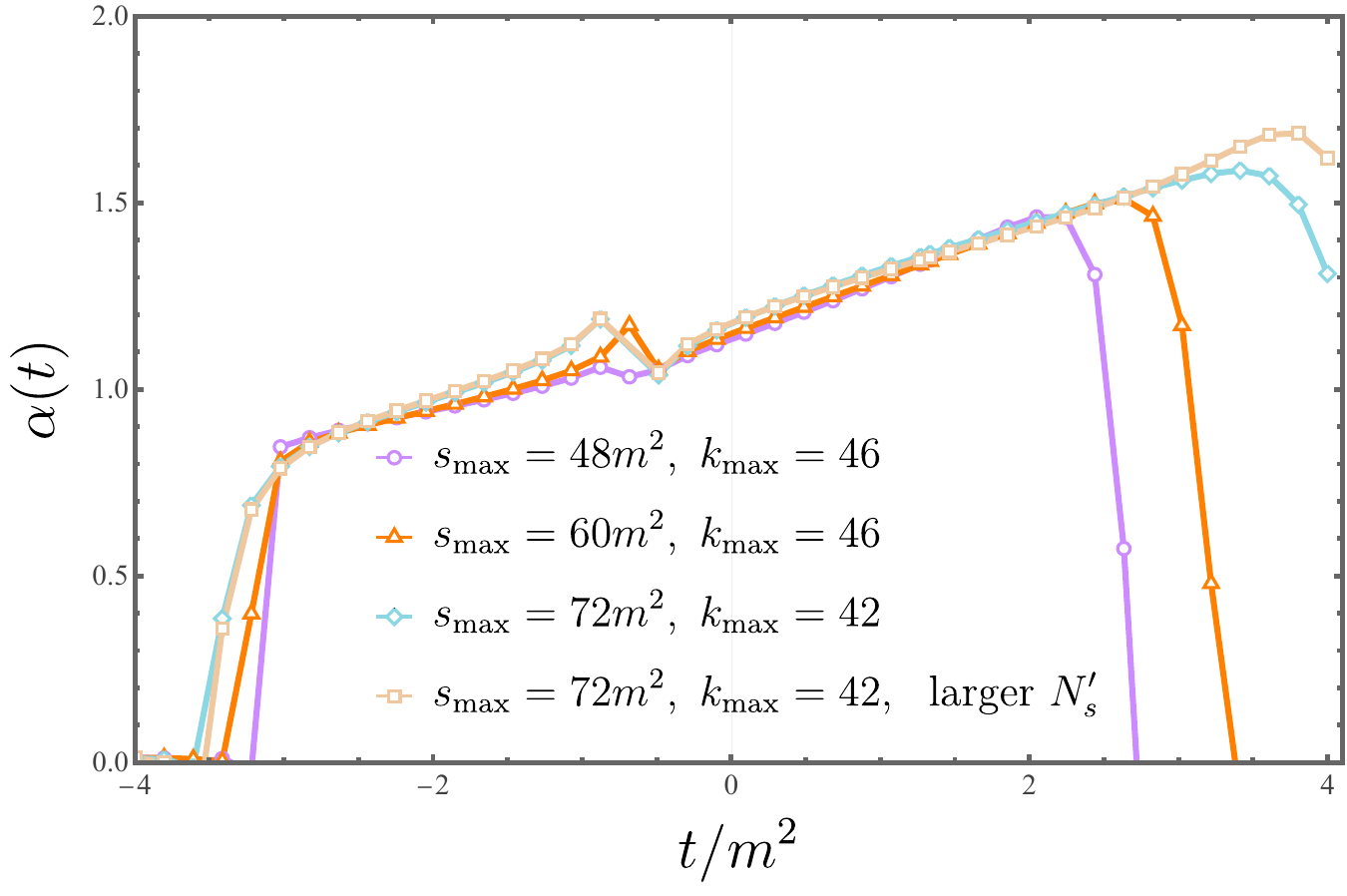}
    \caption{$\alpha(t)$ at ${\rm arg}(s)=\pi/8$ fitted with different numerical parameters $s_{\rm max}$ and $k_{\rm max}$. Numerical parameters are the same as those in Figure \ref{fig:cov_amp}.
    }
    \label{fig:cov_fit}
\end{figure}

Figure \ref{fig:cov_amp} shows the convergence behavior of the fixed-$t$ amplitude at the minimum of $g_0$. We see that, except in the vicinity of $|t| = 4m^2$, it is easy for the amplitude to converge. The poor convergence near $|t| = 4m^2$ occurs because the upper cutoff $s_{\rm max}$ used is not sufficiently large. By increasing $s_{\rm max}$, the amplitude exhibits the expected Regge behavior over a wider range of $t$. This improvement is also reflected in the comparison of $\alpha(t)$ fitted in the range $28m^2$ to $36m^2$, as shown in Figure \ref{fig:cov_fit}. We see that, as $s_{\rm max}$ increases, the (quasi-)linear behavior of $ \alpha(t)$ is pushed to a larger $t$.  Moreover, we find that increasing $N_s'$, {\it i.e.,} imposing linear unitarity for an extended region of $s$, can ensure reliable convergence for larger values of $t$. These observations suggest that $\alpha(t)$ continues to increase as $t$ approaches $4m^2$. Interestingly, even when the amplitudes have not properly converged near $|t| = 4m^2$, the bounds on the amplitude coefficients $g_0$ and $g_2$ have already stabilized, indicating that the region around $t \sim 4m^2$ in the dispersion relation is not crucial for determining the bounds.

\section{Primal bootstrap with parametrized full amplitude}
\label{sec:preprimal}

In this appendix, we briefly review the seminal primal bootstrap method that parametrizes the full amplitude \cite{Paulos:2017fhb}, as we also use this method to probe the Regge behaviors of the amplitude, as a comparison. 

This method relies on the assumption of maximal analyticity, which states that the amplitude (without subthreshold poles) is an analytic function of the Mandelstam variables $s$ and $t$, except along the three branch cuts $s \geq 4m^2$, $t \geq 4m^2$, and $u \geq 4m^2$. By applying Cartan’s B theorem, the amplitude can be analytically continued to a function of $s$, $t$, and $u$ defined on the Cartesian product of three cut planes, with right-hand branch cuts starting at $4m^2$, subject to the constraint $s+t+u = 4m^2$. Note that each of these cut planes can be mapped to a unit disk by the following  variable transformation:
\begin{equation}
\label{rhoDef0}
\rho(x) =\frac{\sqrt{4m^2 - x_0} - \sqrt{4m^2 - x}}{\sqrt{4m^2 - x_0} + \sqrt{4m^2 - x}}\,,
\end{equation}
with the singularities mapped to the boundary of the unit disk. In this paper, we choose $x_0 = 4m^2 / 3$. Therefore, changing the variables from $s$, $t$, and $u$ to $\rho(s)$, $\rho(t)$, and $\rho(u)$ respectively, the amplitude becomes an analytic function on the Cartesian product of three unit disks. One can then perform a triple Taylor expansion around $\rho(s) = \rho(t) = \rho(u) = 0$, which corresponds to $s=t=u=4m^2/3$, to construct a general (full) amplitude:
\begin{equation}
\mc{M}(s,t) = \sum^{'}_{a+b+c \leq N_{\rm max}} d_{abc}\, \rho^{(a}(s) \rho^b(t) \rho^{c)}(u)\,,
\end{equation}
where the $a,b,c$ indices are symmetrized, $d_{abc}$'s are arbitrary coefficients in the ansatz, $N_{\rm max}$ is the truncation order of the expansion, and the prime over the sum indicates that terms redundant under the constraint $s + t + u = 4m^2$ are excluded. (The redundant terms arise because the $s + t + u = 4m^2$ constraints imposes an infinite series of polynomial constraints on $\rho(s)$, $\rho(t)$ and $\rho(u)$.) With this ansatz, which embeds crossing symmetry and analyticity but notably not the Jin-Martin bound, viable amplitudes can then be constructed by projecting onto partial waves and imposing partial wave unitarity conditions.

Note that the unitarity condition permits a divergent behavior in the spin-0 partial wave at the threshold $s = 4m^2$. To account for this, one can add a term exhibiting such divergence,
\begin{equation}\label{eq:th_div}
\alpha_{\rm th}\left(\frac{1}{\rho(s) - 1} + \frac{1}{\rho(t) - 1} + \frac{1}{\rho(u) - 1} \right)\,,
\end{equation}
with $\alpha_{\rm th}$ being an extra parameter, to improve convergence.

\bibliographystyle{JHEP}
\bibliography{ref_new}

\end{document}